\documentclass[aps,prd,superscriptaddress,showpacs,twocolumn,nofootinbib,floatfix,amsmath,amssymb]{revtex4-1}
\usepackage{mathtools}
\usepackage[usenames,dvipsnames]{xcolor}
\usepackage{color}
\usepackage{braket}
\usepackage{bm}
\usepackage{graphicx}
\usepackage{hyperref}

\renewcommand{\vec}[1]{\bm{\mathrm{#1}}}
\def\mat#1{\bm{\mathrm{#1}}}
\def\op#1{\hat{#1}}
\def\opvec#1{\op{\vec{#1}}}

\def\id{I}
\def\tp{\mathrm{T}}
\def\herm{\mathrm{H}}

\def\lcolon{\,\mathopen{:}}
\def\rcolon{\mathclose{:}\,}

\newcommand{\ii}{\mathrm{i}}

\begin{document}

\title{Detectors for probing relativistic quantum physics  beyond perturbation theory}
\author{Eric G. Brown}
\affiliation{Department of Physics and Astronomy, University of Waterloo, Waterloo, Ontario N2L 3G1, Canada}
\author{Eduardo Mart\'{i}n-Mart\'{i}nez}
\affiliation{Department of Physics and Astronomy, University of Waterloo, Waterloo, Ontario N2L 3G1, Canada}
\affiliation{Institute for Quantum Computing and Department of Applied Mathematics, University of Waterloo, 200 University
Avenue W, Waterloo, Ontario, N2L 3G1, Canada}
\affiliation{Perimeter Institute for Theoretical Physics, 31 Caroline St N, Waterloo, ON, N2L 2Y5, Canada}
\author{Nicolas C. Menicucci}
\affiliation{School of Physics, The University of Sydney, Sydney, NSW, 2006, Australia}
 \author{Robert B. Mann}
  \affiliation{Department of Physics and Astronomy, University of Waterloo, Waterloo, Ontario N2L 3G1, Canada}

\begin{abstract}
We develop a general formalism for a non-perturbative treatment of harmonic-oscillator particle detectors in  relativistic quantum field theory using continuous-variables techniques. By means of this we forgo perturbation theory altogether and reduce the complete dynamics to a readily solvable set of first-order, linear differential equations. The formalism applies unchanged to a wide variety of physical setups, including arbitrary detector  trajectories, any number of detectors, arbitrary time-dependent quadratic couplings, arbitrary Gaussian initial states, and a variety of background spacetimes.
As a first set of concrete results, we prove non-perturbatively---and without invoking Bogoliubov transformations---that an accelerated detector in a cavity evolves to a state that is very nearly thermal with a temperature proportional to its acceleration, allowing us to discuss the universality of the Unruh effect. Additionally we  quantitatively analyze the problems of considering single-mode approximations in cavity field theory and show the emergence of causal behaviour  when we include a sufficiently large number of field modes in the analysis. Finally, we analyze how the harmonic particle detector can harvest entanglement from the vacuum. We also study the effect of noise in time dependent problems introduced by suddenly switching on the interaction versus ramping it up slowly (adiabatic activation).
\end{abstract}
\maketitle

\section{Introduction}

For many years, the well-known Unruh-Dewitt model  \cite{DeWitt} has been used to explore aspects of quantum field theory in curved spacetimes.  The great success of this model, which couples a qubit to a quantum field using a simple monopole interaction, has been its use in analyzing the observer dependence of relativistic quantum phenomena. For example, it has provided satisfactory results in the study of  phenomena like the Unruh effect \cite{Unruh1}, meaning that the response of an accelerated qubit detector is thermal with the characteristic Unruh temperature. This result does not require the use of Bogoliubov transformations between inequivalent field expansions and the subsequent tracing over degrees of freedom beyond a horizon. It is instead a consequence of a direct calculation of the response of the detector when traversing a timelike hyperbolic trajectory in spacetime~\cite{BandD}. Additionally, the Unruh-Dewitt model is actually a very good basic description of the light-matter interaction and reproduces quite well the interaction between atoms and light when no exchange of angular momentum is involved~\cite{MigArxiv}.

The main shortcoming of this model is that it is limited to perturbation theory. One is therefore barred from using it to study problems in which a perturbative expansion is not a good approximation. These include strong coupling, long times and high-average-energy exchange processes.

We propose here to model a detector as a quantum harmonic oscillator rather than a qubit, an idea that has been proposed before in other contexts~\cite{UnrhZurek,HuMatacz,MassarSpindel,BeiLok,BerryPh,Ivette}. In other words, we simply replace  two energy levels with infinitely many evenly-spaced levels. Nevertheless, qubits are, in many cases, just approximations to systems with many more levels, so in some ways our description for a particle detector is more natural. Given that most symmetric potentials in nature can be approximated by a harmonic potential for low energies, a harmonic-oscillator detector can model a wide range of detectors, from atomic electromagnetic levels to the molecular vibrational spectrum. In particular, we will consider such detectors in the context of cavity fields (i.e. the fields they interact with will present an IR cutoff), meaning that the field modes are discrete.

Using an oscillator detector has significant advantages over the standard Unruh-DeWitt (qubit-based) detector. First, the quantum evolution can be solved \emph{nonperturbatively}. This results from using the symplectic formalism for Gaussian states and operations~\cite{Schumaker1}. Many of the scenarios of interest in relativistic quantum theory involve quadratic Hamiltonians, making this formalism widely applicable.

Second, the evolution can be evaluated by simply solving (in general numerically) a set of coupled, ordinary, first-order, linear differential equations. Furthermore, the form of this ODE is universal, meaning that one can solve a large range of problems with a very minimal effort. In particular, this approach can be used to solve (a)~arbitrary time-dependent trajectories, (b)~arbitrary quadratic, time-dependent interaction Hamiltonians and boundary conditions, (c)~arbitrary Gaussian initial states of the field modes and detectors, (d)~any number of cavity modes, and (e)~any number of detectors. As we will see, a wide range of different scenarios can therefore be solved non-perturbatively by the same simple differential equation, which implies considerable explanatory power and computational gain.
 For example, as we will see, there is no need to repeat the numeric calculation whenever  we want to change a given initial state if the time-dependent Hamiltonian mediating the interaction is the same. This is rather unlike the perturbative Unruh-DeWitt model in which considerably more effort is required. This universality, plus the ability to sidestep perturbation theory, is the true power of this approach to detector models. We will demonstrate this in Section~\ref{sec:results}.

One obvious limitation of this approach is that to solve the equations in practice, one is forced to apply an infrared cutoff to the field.  However, an infrared cutoff naturally appears when studying quantum field theories in finite volumes (e.g., optical  cavities, periodic waveguides, etc.), and so this formalism enables us to non-perturbatively solve problems of quantum field theories in curved spacetimes inside cavities, a matter of great interest that has not been thoroughly explored to date. If a tabletop experiment in which relativistic quantum phenomena is to appear, discrete systems~\cite{Menicucci2010a} or superconducting circuits \cite{Sab1,PastFutPRL} have an edge on  experimental feasibility.  

Although in  practice one is also forced to use a UV cutoff (namely, computing with only a finite number of modes), in the results presented in this paper we have been careful to find a convergent solution with respect to the number of field modes. Specifically, we run the simulation with more and more modes until the results do not change anymore. As such, this is not a practical limitation. These aspects are especially important in section \ref{secsignaling}, where we study explicitly the effects of an UV cutoff on the causal structure of our setting.

The idea of using harmonic oscillators in relativistic quantum field theory as particle detectors to obtain non-perturbative results was explored by Bei-Lok Hu and collaborators, who reported interesting analytical results in \cite{BeiLok}. Along with its considerable technical accomplishments, this approach emphasized that the Unruh effect is not reliant on gravitational or geometrical arguments, but can be understood as a dynamical effect insofar as it indicates how the quantum vacuum affects the response of a detector contingent on its motion. In general, a detector detects   field quanta with a nonthermal spectrum, where the degree of nonthermality is governed by the parameter that measures the deviation from uniform acceleration \cite{Raval:1996vt}.   However the practical scope of this approach remains to be seen---thus far
it has been limited to very concrete problems in relativistic quantum theory \cite{Lin:2008jj,Hu,Ostapchuk:2011ud}
due to their complexity and the number of assumptions and approximations 
required to obtain quantitative results. 

In performing our analysis we shall employ a more powerful  Gaussian formalism, which provides a more efficient way to address problems of time evolution when considering quadratic Hamiltonian and Gaussian states.  In this sense  our approach is
similar to that of   Dragan and Fuentes  \cite{Ivette}, who made use of the Gaussian formalism to study  a multimode time independent quadratic Hamiltonian of two coupled harmonic oscillators. This approach had some advantages insofar as it did not require any perturbative approximations.  However, their analysis was limited to 1) a single field mode and 2) time-independent Hamiltonian.   Under that proviso, only stationary scenarios and very simple trajectories of detectors can be considered. To study a particular non-inertial scenario (namely eternal uniform acceleration) they relied on the existence of Bogoliubov transformations between inertial modes and Rindler modes, rendering  thermality an a-priori assumption instead of a consequence.  Furthermore, by applying free Bogoliubov transformations to a single field mode, they were unable to see border effects when analyzing the Unruh effect in cavities.  Indeed, the applicability/validity of continuum Bogolibov transformations for eternally accelerating observers  in cavity settings in any regime is a rather obscure topic that has not been thoroughly understood to date.

In what follows,  we present a way to work with an arbitrary time dependent quadratic Hamiltonian and an arbitrary number of modes, being able to  analyze scenarios in cavities for arbitrary trajectories of an arbitrary number of detectors coupled to the field in the cavity without the need to assume any Bogoliubov transformations.  Furthermore, we can overcome causality violation problems \cite{Fay} of  single mode detectors undergoing general trajectories.  We will see that our results are devoid of faster-than-light signalling, unlike previous results limited to single mode approaches.

Far from being a mere presentation of some mathematical tools,  we here obtain non-perturbative answers in very interesting and yet unstudied cavity scenarios:

 (1) \emph{Effects of sudden switching}: It is  known that the response of a detector that is very carefully switched on is very different compared to the interaction being suddenly turned on \cite{Satz}.  We present a brief non-perturbative analysis of this problem showing in a very direct way that it is possible to smoothly switch on the interaction without exciting the detector. 
 
  (2) \emph{Causal signaling inside cavities}: It is well known that imposing a cutoff in the number of field modes allows for acausal signaling \cite{Fay}. This is not surprising since a propagating perturbation in a  cavity cannot be expanded in terms of a finite number of stationary waves. 
 By starting one of the detectors in an excited state and seeing how long it takes the other one to notice its presence  we will demonstrate how causality is recovered in our setting as we increase the number of modes of the cavity. We find the minimum number of cavity modes that must be modelled in order to ensure causality for a given setup.  Note that to completely analyze this kind of process perturbatively would require an analysis up to fourth order.

(3) \emph{The Unruh effect in a cavity}: The question as to what the response of an accelerated particle detector would  be inside a cavity has not been properly explored yet. We will show that the response of an accelerated detector inside a cavity is still thermal, with some corrections coming from boundary effects. Our work provides evidence indicating that the Unruh effect occurs not only for the standard Unruh-DeWitt model, but also for a discretized harmonic oscillator model. Note that we will not rely on any Bogoliubov transformations or quantization process in any accelerated frame. We just answer the question of what the response of an accelerated detector is inside a cavity.  We thereby obtain the Unruh effect inside cavities with a derivation similar to the well-known Unruh-DeWitt result for the continuum of modes \cite{BandD} that additionally  fully sees the appearance of border effects.  Furthermore, we see that the Unruh effect is also present in a completely non-perturbative calculation. The fact that we can easily modify trajectories and interaction types means that we can easily test the model dependence/independence of phenomena like the Unruh effect and entanglement harvesting. Such knowledge is very important for understanding the true physicality of such effects in further research.

(4) \emph{Vacuum entanglement harvesting}: We use the harmonic oscillator detector model to analyze a scenario previously studied in the standard Unruh-Dewitt setup: the extraction of vacuum entanglement by two inertial detectors \cite{Reznik1, VerSteeg2009}. We will also comment on the differences from the harmonic oscillator model and the qubit model in order to see spacelike entanglement.

\section{The Model}

\subsection{Physical setup}

One might suspect that replacing a qubit (two energy levels) with an oscillator (infinite energy levels) in our detector model would complicate the problem, but instead  it \emph{simplifies} it, a fact that has been pointed out previously~\cite{Ivette}. The essential feature that makes this possible is that all states in the problem are Gaussian with no displacement, and all evolutions are homogeneous Gaussian unitaries. This means that all states have Wigner functions that are Gaussian with zero mean, and all evolutions are generated by Hamiltonians that are quadratic in ladder operators with no linear terms. Such Hamiltonians preserve the Gaussian nature of the states~\cite{Schumaker1}. This means that we don't have to keep track of the entire Wigner function; all we need to evolve is the covariance matrix, whose size scales quadratically with the number of modes (rather than exponentially, as is the case for general states). Similarly, all evolutions can be represented by symplectic matrices, which has the same scaling~\cite{Weedbrook2012}.

\begin{figure}[t]
\includegraphics[width=.50 \columnwidth]{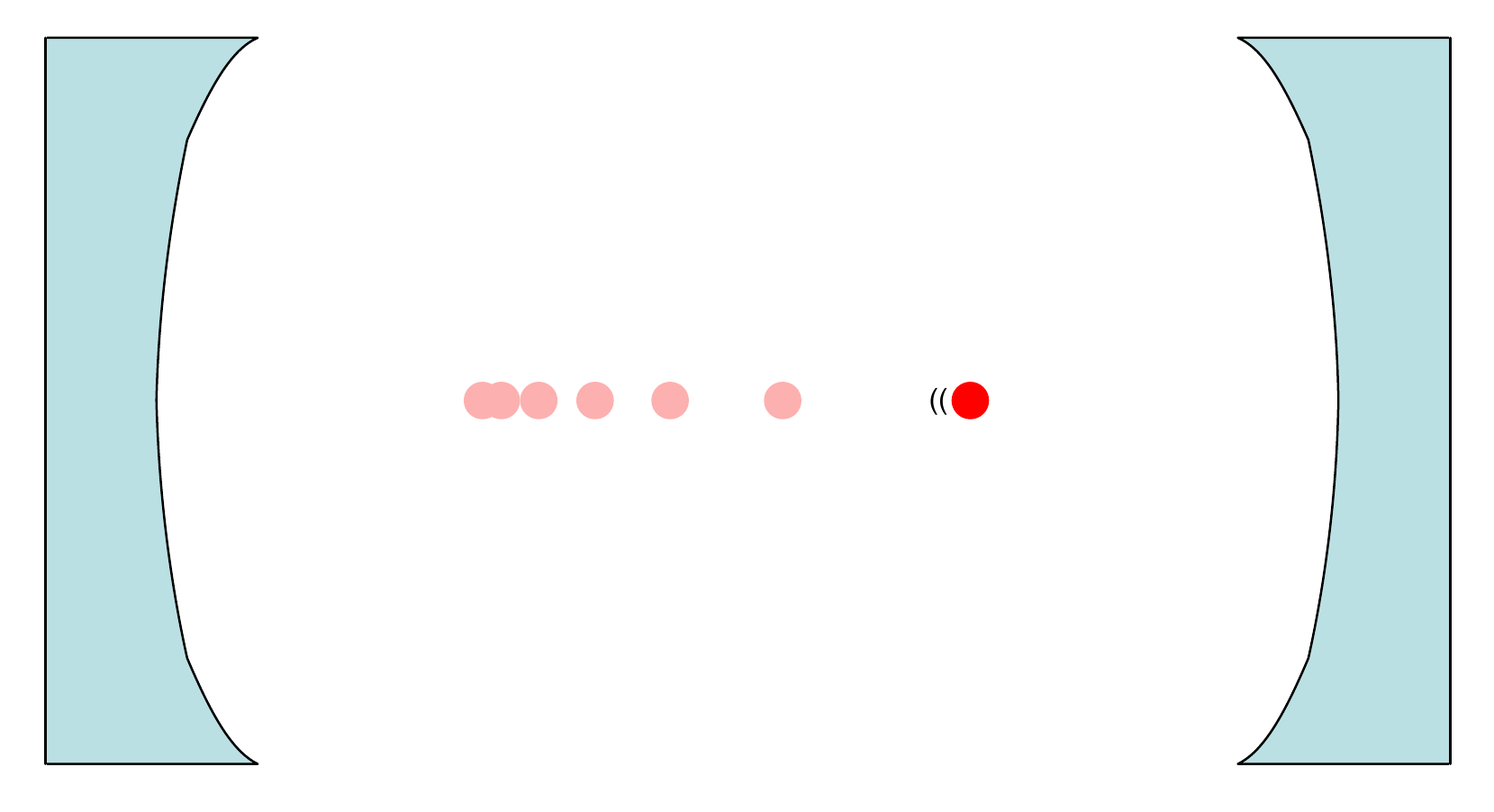}
\caption{Harmonic-oscillator detector (red dot) moving through a fixed cavity. This is to illustrate the difference between our setup and those in which the cavity itself is in motion~\cite{movingcav
}. Note that any number of detectors may be present. This particular case of an accelerated detector is treated in Sec.~\ref{Uther}.}
\label{fig:cavity}
\end{figure}

For calculational purposes, we assume that an IR cutoff of some length $L$ has been imposed on the field.\footnote{This is necessary because we want to use matrix algebra to numerically solve the resulting differential equations, although a formal generalization of our method to the continuum limit may be possible in the form of integro-differential equations resembling those in Section~\ref{sec:solvingevolution}. A complete formulation of this is left to future work.} As such, our physical model is that of a detector moving around in a large cavity. There is an important distinction to be made here with other models that consider the cavity itself as being in motion~\cite{movingcav
}. In our work, by contrast, the cavity is large and fixed, and the detector moves within it (if it moves at all). See Fig.~\ref{fig:cavity}.

The Hamiltonian that we propose below is very similar to one already discussed in the literature in which a (not fully general) Hamiltonian that describes the interaction of a harmonic-oscillator detector with a finite number of field modes was presented \cite{Ivette}. Although their focus was on an approximation of time-independent coupling to just a single field mode (a special case of our approach), the advantages of the Gaussian formalism  and the generality of this approach was made evident. The time-independent single-mode approximation \cite{Ivette} has the advantage of simplifying the obtention of  analytic solutions but at a price of relying on Bogoliubov transformations for uniformly accelerated detectors and limiting to regimes where a single mode approximation can be valid, whereas our results are generally numeric but arise from an approach that has much greater applicability. 
\subsection{Hamiltonians generating evolution with respect to different time parameters}\label{secprel}

In relativistic scenarios it is important to keep in mind that Hamiltonians generate evolution with respect to a given time parameter that does not necessarily coincide with the proper time of some (or any) of the proper times of the physical subsystems that are in interaction. When we consider the global Hamiltonian of multipartite systems we would need to express it in terms of a common time parametrization. Due to this, we also wish to provide a discussion of how to generate evolution with respect to an arbitrary time parameter. In particular it can be in general useful to evolve in a global time coordinate~$t$, particularly in the case where there are multiple detectors in which each detector~$j$ has a different proper time coordinate~$\tau_j$ associated with it. The calculation is most straightforward in the Heisenberg picture, although applies equally well in the Schr\"odinger or interaction pictures. In this way we provide a ``dictionary'' by which we can transform to any other time coordinate. In Sec. \ref{secrel} we continue by introducing the general form of Hamiltonians that can be used with our approach.

Let us proceed, then, by considering a general time-dependent Hamiltonian~$\op H(t)$, which generates translations of the entire system in the global time coordinate~$t$. This Hamiltonian includes the free Hamiltonian for each system, as well as interactions. With respect to~$t$, the Heisenberg equation of motion for a general operator~$\op A(t)$, possibly having explicit dependence on~$t$, is
\begin{align}
\label{eq:Heiseom}
	\frac {d} {dt} \op A(t) &= \frac i \hbar \bigl[\op H(t), \op A(t)\bigr] + \frac {\partial \op A(t)} {\partial t}.
\end{align}
A different choice of time coordinate can be taken into account by applying the chain rule. For the moment, let us choose the new time variable to be the local proper time~$\tau_j$ that parametrizes the worldline~$(x(\tau_j),t(\tau_j))$ traversed by detector~$j$. Applying the chain rule gives
\begin{align}
\label{eq:Heiseomtauj}
	&\frac {d} {d\tau_j} \op A\bigl[t(\tau_j)\bigr] = \left. \frac {dt} {d\tau_j} \frac {d} {dt} \op A(t) \right\rvert_{t=t(\tau_j)} \nonumber \\
	&\qquad= \left. \frac {dt} {d\tau_j} \frac i \hbar \bigl[\op H(t), \op A(t)\bigr] + \frac {dt} {d\tau_j} \frac {\partial \op A(t)} {\partial t} \right\rvert_{t=t(\tau_j)} \nonumber \\
	&\qquad = \frac i \hbar \left[ \left( \frac {dt} {d\tau_j} \op H\bigl[t(\tau_j)\bigr] \right), \op A\bigl[t(\tau_j)\bigr] \right]
	+ \frac {\partial \op A\bigl[t(\tau_j)\bigr]} {\partial \tau_j}\,.
\end{align}
Thus, we can start with the Hamiltonian~$\op H(t)$, which generates translations in the global time coordinate~$t$, and then define
\begin{align}
\label{eq:Httotau}
	\op H_j(\tau_j) \coloneqq \frac {dt} {d\tau_j} \op H\bigl[t(\tau_j) \bigr]
\end{align}
as the Hamiltonian (for the entire system) as seen by detector~$j$, which generates evolution for the entire system with respect to the proper time coordinate~$\tau_j$. The derivative~$dt/d\tau_j$ is the redshift factor for an observer in the detector's reference frame, which provides an overall scaling of all energies in the combined system (because this is what such an observer would experience). Notice that although the notion of proper time is local, we need to be able to evolve the entire system with respect to this coordinate because we are working in the Heisenberg picture. This is not a problem as long as $t(\tau_j)$ is an invertible function over the range of times of interest. The equivalence of the two pictures is made explicit by defining Heisenberg operators that are more natural to the detector's frame:
\begin{align}
\label{eq:Ajoftauj}
	\op A_j(\tau_j) \coloneqq \op A\bigl[ t(\tau_j) \bigr]\,.
\end{align}
We can now use Eqs.~\eqref{eq:Httotau} and~\eqref{eq:Ajoftauj} to rewrite Eq.~\eqref{eq:Heiseomtauj} as
\begin{align}
	\frac {d} {d\tau_j} \op A_j(\tau_j) &= \frac i \hbar \bigl[\op H_j(\tau_j), \op A_j(\tau_j) \bigr] + \frac {\partial \op A_j(\tau_j)} {\partial \tau_j}\,.
\end{align}
For example, if $\op A(t)$ were to represent the position of the second hand on a wristwatch worn by an observer traveling with detector~$j$, then it would make more sense to consider $\op A_j(\tau_j)$ because this operator would have a simpler evolution with respect to~$\tau_j$ than $\op A(t)$ would with respect to~$t$ (since the wristwatch evolves more simply with respect to~$\tau_j$ than with respect to~$t$). Similarly, it will be easier to start with the simple version of the wristwatch's Hamiltonian~$\op H_j(\tau_j)$ and then invert Eq.~\eqref{eq:Httotau} to obtain
\begin{align}
\label{eq:Htautot}
	\op H(t) = \frac {d\tau_j} {dt} \op H_j\bigl[\tau_j(t) \bigr]
\end{align}
(which will be more complicated) for use in the global Hamiltonian.

The upshot of all of this is that we can define a single Hamiltonian~$\op H(t)$ for the whole system with respect to some global time coordinate~$t$ and then use Eq.~\eqref{eq:Httotau} to transform it to any other time coordinate we wish to use for the evolution. Furthermore, when building up this Hamiltonian, it will sometimes be easier to start by defining a piece of it with respect to local proper time and then use Eq.~\eqref{eq:Htautot} to figure out what this piece looks like in the global time coordinate.

\subsection{The Unruh-DeWitt Hamiltonian in general scenarios}\label{secrel}

We have to be careful when we want to deal with Hamiltonians generating translations with respect to different time parameters, above all when we want to describe the interaction of systems that have different proper times.

Indeed, in general scenarios it is not trivial to define either the interaction nor the free Hamiltonian in different pictures. To guide the reader through this section let us introduce the following notation: We will call $\op H^\text{S},\op H^\text{D},\op H^\text{H}$ respectively the complete Hamiltonian in the Schr\"odinger, interaction (Dirac) and Heisenberg pictures. We will denote with the subscript `0' the free part of the Hamiltonian and `1' the interaction part. Also, we will include a superindex $t$ or $\tau$ denoting with respect to which time the Hamiltonian is a generator of translations.

We will consider the interaction of a number of particle detectors with a quantum field. To model this interaction we will consider an X-X coupling of the form of the Unruh-DeWitt Hamiltonian \cite{DeWitt}. Note, however, that the formalism we present is much more general than this, and we can in fact use any quadratic Hamiltonian that we like. For our immediate purposes we choose to use the X-X coupling in order to compare with previous works. 

For every detector coupled to the field, the Unruh-DeWitt interaction will be described by the following Hamiltonian
\begin{equation}
H_{1}= \lambda(\tau) \op \mu\, \op \phi[x(\tau)],
\end{equation}
where $\lambda(\tau)$ is the switching function, $\op \mu$ is the monopole moment of the detector and $\op\phi[x(\tau)]$ is the field operator evaluated along the worldline of the detector parametrized in terms of the time $\tau$ with respect to which the Hamiltonian generates translations .

For the sake of clarity, let us start our reasoning with a very simple scenario: let us consider a single detector undergoing general motion in flat spacetime with an associated proper time $\tau$ and a scalar quantum field that we will choose to expand in terms of plane-wave solutions in terms of a global Minkowskian time $t$, as is commonplace in quantum field theory.

To derive the correct form of the Hamiltonian in the interaction and Heisenberg pictures let us first write the field and monopole operators in the Schr\"odinger picture:
\begin{align}
\op \phi^{\text{S}}[x(\tau)] &= \sum_n\!\Big(\op a_n v_n[x(\tau)] +\op a^\dag_n v_n[x(\tau)] \Big), \\
\op \mu^{\text{S}} &= (\op a_d+\op a^\dag_d),
\end{align}
where $v_n(x)$ are the spatial part of the solutions to the field equations, i.e. the mode functions. For instance, in the case of reflecting boundary conditions these would be $v_n(x)=\sin(k_nx)$, whereas in the case of periodic ones $v_n(x)=e^{\ii k_nx}$. Here $k_n$ is the wavevector for field mode $n$; we will specify its form below.

Let us start from a very well known result from first principles: we can write  the the free Hamiltonian for the field, and the free Hamiltonian of the detector in their respective times in the Schr\"odinger picture:
\begin{align}
\op H_{0,\text{field}}^{\text{S},t}&=\sum_n\omega_n \op a_n^\dag \op a_n\,, \\
\op H_{0,\text{det}}^{\text{S},\tau}&=\Omega \op a_{d}^\dag \op a_d\,.
\end{align}

Now, to write the complete free Hamiltonian we cannot just naively sum these two terms together because they generate translations with respect to different time parameters. We would need first to transform them to a common time parameterization. We will see that in order to recover the correct form of the well-known Unruh-Dewitt Hamiltonian in the interaction picture, we must transform the field Hamiltonian to generate translations in the proper time of the detectors. In this way, using \eqref{eq:Htautot} we have that
\begin{align}
\op H_{0,\text{field}}^{\text{S},\tau}=\frac{d}{d\tau}t(\tau)\sum_n\omega_n \op a_n^\dag \op a_n \,,
\end{align}
so that we can write the complete Hamiltonian in the Schr\"odinger picture generating translations in $\tau$ as 
\[\op H^{\text{S},\tau}=\op H_{0}^{\text{S},\tau}+H^{\text{S},\tau}_{1},\]
where
\begin{align}
\op H_{0}^{\text{S},\tau}&=\frac{d}{d\tau}t(\tau)\sum_n\omega_n \op a_n^\dag \op a_n 
+\Omega \op a_{d}^\dag \op a_d\\
\op H_{1}^{\text{S},\tau}&=\lambda(\tau)(\op a_d+\op a^\dag_d)\sum_n\!\Big(\op a_n v_n[x(\tau)] +\op a^\dag_n v_n[x(\tau)] \Big)
\end{align}
Note that the free Hamiltonian is not time independent as it was in the case where the detector is inertial. 

In most textbooks \cite{BandD}, calculations involving non-inertial detectors coupled to the field are dealt with in the interaction picture.  We will see that we recover the well-known form of the interaction Unruh-DeWitt Hamiltonian by changing from the Schr\"odinger to the interaction picture.

Recall the transformation between the Schr\"odinger and the Interaction pictures:
\begin{align}
\op H^{\text{D},\tau}&=\op U_0^\dagger(\tau) \op H^{\text{S},\tau} \op U_0(\tau)\,,
\end{align}
where $\op U_0(\tau)$ is the solution to the Schr\"odinger equation in~$\tau$ using just the free Hamiltonian~$\op H_0^{\text{S},\tau}$:
\begin{align}
\label{eq:SchrodU0}
	i \frac {d} {d\tau} \op U_0(\tau) = \op H_0^{\text{S},\tau} \op U_0(\tau)\,,
\end{align}
Notice that in this case the transformation is non-trivial due to the non-trivial dependence on $\tau$ of the global time parameter $t(\tau)$. This yields an  explicit time dependence of the field's free Hamiltonian. Since $\op H_0^{\text{S},\tau}$ commutes with itself at different times, we can solve Eq.~\eqref{eq:SchrodU0} explicitly without needing to worry about time ordering:
\begin{align}
\op U_0(\tau) &= \exp\left[ -\ii\int_0^\tau d\tau\, \op H_0^\text{S} \right] \nonumber \\
&=\exp\Big[-\ii \int_0^{\tau} d\tau \Big(\frac{d[t(\tau)]}{d\tau}\sum_n\omega_n \op a_n^\dag \op a_n 
+\Omega \op a_d^\dag \op a_d\Big)\Big] \nonumber \\
&=\exp\Big[-\ii  \Big(\sum_n\omega_n \op a_n^\dag \op a_n\Big) t(\tau)
- \ii \Omega \op a_d^\dag \op a_d\tau\Big],
\end{align}
This operator leaves invariant the free parts of the Hamiltonian, and its action on the $\op a_n$ and $\op a_{d_j}$ operators is
\begin{align}
\op U_0^\dag(\tau) \op a_n \op U_0(\tau)&=e^{-\ii \omega_n t(\tau)} \op a_n \,, \\
\op U_0^\dag(\tau) \op a_{d} \op U_0(\tau)&=e^{-\ii \Omega \tau} \op a_{d}\,,
\end{align}
allowing us to write the Unruh-Dewitt Hamiltonian in the interaction picture with respect to the parameter $\tau$ as
 \begin{align}\label{hamilgood}
\op H^{\text{D},\tau}&=\frac{dt(\tau)}{d\tau}\sum_n\omega_n \op a_n^\dag \op a_n 
+\Omega \op a_{d}^\dag \op a_d+(\op a_d e^{-\ii\Omega\tau}+\op a^\dag_d e^{\ii\Omega\tau}) \nonumber \\
&\times \lambda(\tau)  \sum_n\!\Big(\op a_n u_n[x(\tau),t(\tau)] +\op a^\dag_n u^*_n[x(\tau),t(\tau)] \Big)\,,
\end{align}
where
\begin{equation}
 u_n[x(\tau),t(\tau)]= e^{-\ii\omega_n t(\tau)}v_n[x(\tau)]\,.
\end{equation}
Thus we recover the standard form of the Unruh-Dewitt Hamiltonian \cite{BandD} in the interaction picture that generates translations with respect to time $\tau$ starting from  the well known free Hamiltonians (in the Schr\"odinger pictures, with respect to their respective natural time parameters) after transforming to a common time $\tau$ and changing to the interaction picture.

Notice that the (real-valued) coupling parameters~$\lambda(\tau)$ can be time dependent. This allows for, among other things, switching the detector on and off with a particular temporal profile known as the switching function or time window function. The last pair of terms indicates that the detector interacts with the field with a strength (and phase) governed by the mode functions~$u_n(x,t)$. In order to determine these mode functions, we need to establish boundary conditions for the cavity. These can include, for example, Dirichlet boundary conditions, in which the field strength vanishes at the boundary (as in a physically realistic optical cavity) or periodic conditions (as in a physically realistic periodic waveguide). The mode functions for these cases take the forms
\begin{align}
	u_n(x,t) &\xmapsto{\text{Dirichlet}} \exp \left(-\ii \omega_n  t \right) \sin \left( k_n x \right) \label{modefunction1}\\
\intertext{and}
	u_n(x,t) &\xmapsto{\text{periodic}} \exp \left(-i \omega_n t \right) \exp \left( i k_n x \right)\label{modefunction2}\,,
\end{align}
where, as we are going to work with massless fields, $\omega_n=|k_n|$ and $k_n=n\pi/L$ or $k_n=2n\pi/L$ respectively for the Dirichlet or periodic boundary conditions.

Now, we will find it convenient for use in the next section to work in the Heisenberg picture. We will use the fact that the form of the complete Hamiltonian in the Heisenberg picture coincides with the form of the Hamiltonian in the Schr\"odinger picture. To see this, note that the transformation between the two pictures is by the full time-evolution operator~$\op U(\tau)$, which satisfies the full Schr\"odinger equation
\begin{align}
\label{eq:SchrodU}
	i \frac {d} {d\tau} \op U(\tau) = \op H^{\text{S},\tau} \op U(\tau)\,.
\end{align}
The Hamiltonian in the Heisenberg picture is obtained from its Schr\"odinger-picture counterpart by the usual transformation between the two pictures for any operator:
\begin{align}
\label{transfo}
	\op H^{\text{H},\tau} &= \op U(\tau)^\dag \op H^{\text{S},\tau} \op U(\tau)\,.
\end{align}
This means that we do not have to do any work to modify the Schr\"odinger-picture Hamiltonian in order to use it in the Heisenberg picture. All we have to do is reinterpret all operators within it as being Heisenberg-picture operators instead of Schr\"odinger-picture ones.

After all these simple steps, it is straightforward to write the most general X-X type Hamiltonian for an arbitrary number of detectors undergoing general trajectories with different proper times $\tau_j$ and with time dependent couplings. However, if multiple detectors have different proper times then we again need to be careful. One must always make a choice of time, and in this more general case it makes sense to use the global Minkowski time $t$. Transforming the Hamiltonian to time $t$ and reinterpreting all operators in the Heisenberg picture yields
\begin{align}\label{eq:hamilto}
\nonumber \op H^{\text{H},t}&=\sum_{n=1}^N \omega_n \op a_n^\dag \op a_n 
+\sum_{j=1}^M \frac{d\tau_j(t)}{dt}\Big[\Omega_j \op a_{d_j}^\dag \op a_{d_j}\\
&+\sum_{n=1}^N \lambda_{nj}(t)(\op a_{d_j}+\op a^\dag_{d_j})\big(\op a_n v_n[x_j(t)] +\op a^\dag_n v_n[x_j(t)] \big)\Big]
\end{align}
where $x_j(t)$ is the trajectory of the $j$-th detector parametrized in terms of the global Minkowskian time $t$, and all operators are now understood to be in their Heisenberg representation. We will see in the next section how working in this representation allows us to derive a simple, number-valued equation of motion that describes the full evolution of the detectors+field state.

To recapitulate, the Hamiltonian above represents a set of ${N+M}$ time-dependent coupled harmonic oscillators. $M$ of them are oscillator-based Unruh-DeWitt detector modes (labeled with the index $d_j$), and the other $N$ are modes of the quantum field inside a cavity (labeled with an integer index~$n$). Notice that the there is no direct detector-detector coupling, and the field is a free field, meaning there is no coupling directly between field modes either.

We emphasize that Eq.~\eqref{eq:hamilto} is not the most general form of the Hamiltonian that could be imagined in this scenario. It is simply the same as the original Unruh-DeWitt detector model. The connection is made by choosing no relative phase between $\op a_{d_j}$ and $\op a_{d_j}^\dag$ in the interaction term, which makes their sum proportional to the position (monopole moment) of the oscillator. In general, our formalism, to be presented now, is capable of solving far more general interaction models, provided they are quadratic.

Our problem is now this: given detector worldlines $\bigl[ t(\tau_j),x(\tau_j) \bigr]$ and an initial (Gaussian) state for the detectors and field, evolve the detectors and the field using the Hamiltonian in Eq.~\eqref{eq:hamilto}, and consider the reduced state of the detectors after the evolution. In order to make use of the simplification afforded by the use of Gaussian states and quadratic Hamiltonians, in the next section we will derive a differential equation using the symplectic formalism for the Heisenberg picture and another differential equation using the Hilbert space evolution in the Interaction picture.  Working with either of these pictures will let us compute the covariance matrix for the field and detectors throughout the evolution, and since the state remains Gaussian the whole time, this is equivalent to tracking the evolution of the full state itself.

\section{Solving the Evolution}\label{sec:solvingevolution}
%

In the case of an oscillator detector the great advantage is that one is able to utilize the Gaussian formalism \cite{Schumaker1,Adesso1,Menicucci2011,Weedbrook2012}. In this solution method we will find it convenient to use the Heisenberg picture. As discussed in Section~\ref{secrel}, there are subtleties involved because of the different time coordinates available (proper time for each detector, plus the global time coordinate of the field).

In section \ref{phase} we will use the Heisenberg-picture Hamiltonian, which generates translations in the global coordinate time~$t$, to derive a simple linear differential equation using the symplectic formalism for Gaussian evolutions. This is the simplest, most straightforward, and computationally most efficient approach.

In section \ref{hilbert}, we describe another method that directly calculates using the interaction Hamiltonian in the interaction picture. This method is more complicated than the symplectic method and yields a non-linear system of ODEs, but because it is independently derived, and because the resulting differential equations are completely different (second-order instead of first-order), this provides an independent check of our numerical results.

The purpose of presenting in parallel both approaches is dual: on the one hand we can tackle problems with two different approaches (which provides a way to optimize our computational strategy in order to solve a given problem). On the other hand we see that the same results can be obtained via independent methods. This serves as a connection between the two formalisms, and a consistency check.

\subsection{Phase-space evolution} \label{phase}

We will use the Hamiltonian \eqref{eq:hamilto} in the Heisenberg picture to evolve Heisenberg quadrature operators~$(\op q_{d_j}(t), \op p_{d_j}(t))$ for each detector and $(\op q_n(t), \op p_n(t))$ for each field mode. To streamline calculations, we stack these operators on top of each other to form the following vector of operators, while omitting the explicit $t$-dependence for clarity:
\begin{equation}
\opvec x \coloneqq \left(\op q_{d_1},\dots,\op q_{d_M}, \op q_{1},\dots,\op q_{N},\op p_{d_1},\dots,\op p_{d_M},\op p_{1},\dots,\op p_N\right)^\tp\,,
\end{equation}
where
\begin{equation}
 \op q_i=\frac{1}{\sqrt{2}}\big(\op a_i+\op a_i^\dag\big),\quad \op p_i=\frac{\ii}{\sqrt{2}}\big(\op a_i^\dag-\op a_i\big)
\end{equation}
are respectively the canonical position and momentum of every single oscillator in the Heisenberg picture. Note that the transpose operation~$^\tp$ merely transposes the shape of an operator-valued vector; it does nothing to the operators themselves.

A Gaussian state is one that can be expressed purely in terms of its first and second quadrature moments. If we neglect phase-space displacements, this means that such a state is fully characterized by a covariance matrix~$\mat \sigma$, the entries of which are
\begin{align}
	\sigma_{i j} \equiv \langle \op{x}_i \op{x}_j + \op{x}_j \op{x}_i \rangle - 2\langle \op{x}_i \rangle \langle \op{x}_j \rangle\,.
\end{align}
Specifically, the Wigner function for the zero-mean Gaussian state associated with~$\mat \sigma$ is
\begin{align}
\label{eq:Wignersigma}
	W(\vec x) = \pi^{-(N+M)} (\det \mat\sigma)^{-1/2} \exp(-\vec x^\tp {\mat\sigma}^{-1} \vec x)\,,
\end{align}
where $\vec x$ is the vector of c-number-valued coordinates corresponding to $\opvec x$:
\begin{equation}
\vec x \coloneqq \left( q_{d_1},\dots, q_{d_M},  q_{1},\dots, q_{N}, p_{d_1},\dots, p_{d_M}, p_{1},\dots, p_N\right)^\tp\,,
\end{equation}
Although in general, displacements will be required for a full description of Gaussian states and evolution, if we start with a zero-mean initial field state (such as the Minkowski vacuum) and a zero-mean initial detector state as well, then the lack of linear terms in our Hamiltonian means that the state remains at zero mean at all times.

To compute the evolution, we utilize the fact that quadratic Hamiltonians preserve Gaussianity~\cite{Schumaker1}. This means that quadrature operators get mapped to linear combinations of quadrature operators:
\begin{align}
\label{eq:Heisenbergsymplectic}
	\opvec x' = \op U^\dag \opvec x \op U = \mat S \opvec x\,,
\end{align}
where $\op U$ is a Gaussian unitary (i.e.,~a unitary transformation generated by a quadratic Hamiltonian). In this equation, $\mat S$ is a symplectic matrix of c-numbers that acts via matrix multiplication on $\opvec x$ as a vector, while $\op U$ is a unitary operator that acts on the individual operators within~$\opvec x$.  Specifically,
\begin{align}\label{eq:specific_symplectic_trafo}
	\op x_j' = \op U^\dag \op x_j \op U = \sum_{k=1}^{2(N+M)} S_{jk} \op x_k\,.
\end{align}
Notice that in general there would be a phase-space displacement term, which would give $\opvec x' = \mat S \opvec x + \vec y$, but we are neglecting this as justified above. The symplectic nature of~$\mat S$ is guaranteed because the commutation relations must be preserved, giving rise to a symplectic form~$\mat \Omega$ to be preserved by the Heisenberg matrix action.  The explicit form of~$\mat \Omega$ may be deduced by writing out the commutation relations for $\opvec x$ and requiring them to be unchanged under the Gaussian unitary operation.  Using the notation of Ref.~\cite{Menicucci2011},  the canonical commutation relations~$[\op q_j, \op p_k] = i \delta_{jk}$ (with $\hbar = 1$) can be written succinctly as
\begin{align}
\label{eq:xcomm}
	[ \opvec x, \opvec x^\tp ] = i
	\begin{pmatrix}
		\mat 0	& \mat \id	\\
		-\mat \id	& \mat 0
	\end{pmatrix}
	=: i \mat \Omega\,,
\end{align}
where the commutator of two operator-valued vectors is defined as
\begin{align}
\label{eq:commdef}
	[ \opvec r, \opvec s^\tp ] := \opvec r \opvec s^\tp - (\opvec s \opvec r^\tp)^\tp\,.
\end{align}
Therefore, $\mat \Omega$ has elements $\Omega_{ij}=-i \big[\op{x}_i,\op{x}_j \big]$. That $\mat S$ is symplectic means that $\mat S \mat \Omega \mat S^\tp = \mat \Omega$, which follows from requiring that the new commutation relations must equal the old ones: $[\mat S \opvec x, (\mat S \opvec x)^\tp] = i\mat \Omega$. (See Ref.~\cite{Menicucci2011} for more details.)

A general quadratic Hamiltonian generates a Gaussian unitary~$\op U(t) = e^{-i\int \op H^{\text{H},t} dt}$ that is associated, by Eq.~\eqref{eq:Heisenbergsymplectic}, with a symplectic matrix~$\mat S(t)$, which satisfies
\begin{align}
\label{eq:xevol}
	\opvec x(t) = \op U(t)^\dag \opvec x_0 \op U(t) = \mat S(t) \opvec x_0\,,
\end{align}
where all time dependence (or lack thereof) is indicated explicitly, and $\opvec x_0$ is the initial vector of quadratures at ${t=0}$. Correspondingly, the Schr\"odinger evolution of the state, as given by the evolution of the covariance matrix, takes the form
\begin{align}
	\mat \sigma(t)= \mat S(t)\mat \sigma_0 \mat S(t)^\tp,
\end{align}
where $\mat \sigma_0$ is the initial state. Our goal in this section is thus to find a differential equation for $\mat S(t)$ that represents the evolution generated by a quadratic Hamiltonian.

A general time-dependent, quadratic, Heisenberg-picture Hamiltonian $H$ with generated time coordinate~$t$ can be written as
\begin{equation}\label{eq:canon}
	\op H=\opvec x^\tp\mat F(t)\opvec x\,,
\end{equation}
where $\mat F(t)$ is a Hermitian matrix of c-numbers containing any explicit time-dependence of the Hamiltonian. We can now write the Heisenberg equation for the time evolution of the quadratures:
\begin{equation}
\label{eq:timederivx}
	\frac{d}{dt}\opvec x=i\big[\op H,\opvec x\big]
\end{equation}
Writing this out in components and using Eqs.~\eqref{eq:xcomm} and~\eqref{eq:canon} gives
\begin{align}
\label{eq:timederivxcomp}
	\frac{d}{dt}\op x_j &= i\big[\op H,\op x_j \big] = i\sum_{mn}{F}_{mn}(t)\big[\op x_m \op x_n,\op x_j\big] \nonumber \\
	&= \sum_{mn}{F}_{mn}(t) \Bigl( \op x_m \Omega_{jn} + \Omega_{jm} \op x_n \Bigr)\,,
\end{align}
which can be collected back into vector form as
\begin{align}
	\frac{d}{dt}\opvec x 
	&= \mat \Omega \mat F^{\text{sym}}(t) \opvec x\,, 
\end{align}
where $\mat F^{\text{sym}}=(\mat F+\mat F^\tp)$. We now plug in Eq.~\eqref{eq:xevol}, giving
\begin{equation}\label{step4}
	\frac{d}{dt}\big[\mat S(t)\big] \opvec x_0=\mat \Omega \mat F^{\text{sym}}(t)\mat S(t)\opvec x _0\,,
\end{equation}
where we used the fact that $\opvec x _0$ is time independent. Now we use Eqs.~\eqref{eq:xcomm} and~\eqref{eq:commdef} to eliminate the operators by taking commutators with $\opvec x_0^\tp$ on both sides:
\begin{align}
\label{eq:xevolcomm}
	\left[ \left(\frac{d}{dt}\mat S(t) \right)\opvec x_0, \opvec x_0^\tp \right] &= \Bigl[ \mat \Omega \mat F^{\text{sym}}(t)\mat S(t)\opvec x _0, \opvec x_0^\tp \Bigr] \nonumber \\
	\frac{d}{dt}\mat S(t) [ \opvec x_0, \opvec x_0^\tp ] &= \mat \Omega \mat F^{\text{sym}}(t)\mat S(t) [ \opvec x _0, \opvec x_0^\tp ]\,,
\end{align}
where we have factored out the c-number matrices multiplying $\opvec x_0$~\cite{Menicucci2011}. Since $[ \opvec x_0, \opvec x_0^\tp ] = i\mat \Omega$, which is an invertible matrix of c-numbers, we can cancel it, yielding the following first-order, linear, ordinary differential equation for the symplectic matrix:
\begin{equation}\label{step4}
\frac{d}{dt}\mat S(t)=\mat \Omega \mat F^{\text{sym}}(t)\mat S(t).
\end{equation}
Solving this equation with the initial condition $\mat S(0)=\mat I$ such that $\opvec x_0=\mat S(0)\opvec x_0$ is completely equivalent to solving the standard Hilbert space evolution with the Hamiltonian-unitary formalism after taking advantage of the quadratic nature of the Hamiltonian, as we will see in the next section. 

It will be convenient to know the form of the Hamiltonian in terms of the annihilation and creation operators of the system, since this is how the monopole-monopole coupling is typically given. To this end, let us stack ladder operators on top of each other to form the following column vectors:
\begin{align}\label{avector}
	\opvec a &\coloneqq (\op a_{d_1},\dots,\op a_{d_M}, \op a_1,\dots \op a_N)^\tp, \nonumber \\
	\opvec a^\dag &\coloneqq (\op a_{d_1}^\dag,\dots,\op a_{d_M}^\dag, \op a_1^\dag,\dots \op a_N^\dag)^\tp. 
\end{align}
The Hamiltonian from Eq.~\eqref{eq:canon} can be put into the form
\begin{align} \label{hamil2a}
	\op H = (\opvec a^\dag)^\tp \mat w(t) \opvec a +(\opvec a^\dag)^\tp \mat g(t) \opvec a^{\dag} +\opvec a^\tp \mat g(t)^\herm \opvec a\,,
\end{align}
where $\mat w(t)$ and $\mat g(t)$ are coefficient matrices, and $\mat M^\herm = \mat M^{*\tp} = \mat M^{\tp*}$ denotes the conjugate transpose of any matrix $\mat M$.

By equating \eqref{hamil2a} to \eqref{eq:canon} and comparing coefficients it is easy to obtain the form of the matrix $\mat F(t)$, which takes the following block form:
\begin{equation}
\mat F(t)=\left(
\begin{array}{cc}
\mat A(t) & \mat X(t)\\
\mat X(t)^\herm& \mat B(t)
\end{array}\right)\,,
\end{equation}
where
\begin{align}
\mat A(t)&=\frac12\left(\mat w(t)+\mat g(t)+\mat g(t)^\herm\right)\,,\\
\mat B(t)&=\frac12\left(\mat w(t)-\mat g(t)-\mat g(t)^\herm\right)\,,\\
\mat X(t)&=\frac i 2\left(\mat w(t)-\mat g(t)+\mat g(t)^\herm\right)\,.
\end{align}

Note that  in the simple particular cases where the Hamiltonian at different times commute the solution of \eqref{step4} can be obtained analytically and it is simply $\mat S(t)=\exp(\mat \Omega \mat F^{\text{sym}}t)$. 

As a final note, we would like to point out that we derived these results using global Minkowski time~$t$ because it is the least biased time coordinate when there are multiple detectors involved. But any time coordinate can be used instead---such as some particular detector's proper time---as long as the appropriate transformation is made to the Hamiltonian via Eq.~\eqref{eq:Httotau}. Doing so would define a new differential equation of the same form as Eq.~\eqref{step4} but evolving in the new time coordinate instead of in~$t$. In fact, this is the most direct way to calculate detector responses, and it is the method that we use in Section~\ref{sec:results}.

\subsection{Hilbert-space evolution} \label{hilbert}

Although the formalism presented in the section above is elegant and fully general for quadratic Hamiltonians, it is useful to compare our results with an independent method. This section outlines the supplementary method that we used for this purpose.
Without invoking the full machinery of symplectic transformations, we can still take advantage of the quadratic nature of the interaction to derive the same results by numerically calculating the unitary time evolution operator directly in the interaction picture, an approach that is more standard within quantum field theory. Here we give only an outline of the method;  details can be found in Appendix~\ref{appendixA}.

For a quadratic Hamiltonian, time evolution can be expressed in terms of displacements, squeezing, and rotation unitary operations \cite{Ma1}. In particular, the unitary evolution we are looking to solve for can then be put into the form
\begin{align} \label{combo}
	\op U(\tau)=e^{i\gamma(t)}\op S(\mat z(t))\op D(\vec \beta(t))\op R(\mat \phi(t))\,,
\end{align}
where $\gamma$ is number valued and the squeezing, rotation and displacement operators are respectively defined as
\begin{align} \label{squeeze}
	\op S(\mat z) &= e^{ \frac{1}{2} \big[(\opvec a^\dag)^\tp  \mat z \opvec a^\dag- \opvec a^\tp\mat z^\text{H} {\opvec a}\big] }, \\ \label{rotation}
	\op R(\mat \phi)&= e^{i (\opvec a^\dag)^\tp {\mat \phi} {\opvec a}}, \\ 
	\op D(\vec \beta)&= e^{\vec \beta^\tp \opvec a^\dag -\vec \beta^\herm \opvec a},
\end{align}
where $\mat z$ and $\mat \phi$ are matrices, and $\vec \beta$ is a column vector. The notation $\mat z^\text{H}$ is used to represent the  conjugate transpose of $\mat z$, the elements of which are number valued. Note that without loss of generality we can consider $\mat z$ to be symmetric because it is only the symmetric part that contributes to $\op{S}(\mat z)$. Also note that $\mat \phi$ must be a Hermitian matrix $(\vec \phi =\vec \phi^{\text{H}})$ to ensure unitarity of $\op{R}(\mat \phi)$.

The exact form of these transformations can be obtained nonperturbatively by employing a technique introduced by Heffner and Louisell~\cite{Heffner1}. For the cases of interests (for the family of Hamiltonians \eqref{hamilgood}), and as it is thoroughly detailed in  appendix \ref{appendixA}, 
due to the algebraic nature of the interaction, the evolution can be reduced to
\begin{align} \label{soln2}
	\op{U}=e^{i\gamma}\op{S}(\mat z)\op{R}(\mat \phi),
\end{align}
and the problem of solving the dynamics can then be reduced to solving the following system of coupled differential equations
\begin{align}\label{caponata}
&i\dot{\mat C}(t)=4\mat C_s(t) \mat g^\text{H}(t) \mat C_s(t)+2\mat w(t) \mat C_s(t)+\mat g(t),\\
&i \dot{\mat D}(t)=(4\mat C_s(t) \mat g^\text{H}(t)+\mat w(t))(\mat D(t)+\mat I), \label{cefalina}
\end{align}
where $\mat w(t)$ and $\mat g(t)$ are the Hamiltonian coefficient matrices as defined in Eq. (\ref{hamil2a}), evaluated in any picture one chooses (for example this can be done directly with the interaction Hamiltonian in the interaction picture). We have defined $\mat C_s \equiv (\mat C+\mat C^\text{T})/2$, where the matrices $\mat C_s$ and $\mat D$ are identified with the squeezing and rotation by 
\begin{align}
	\mat C_s=\frac{1}{2}\tanh(\mat r)e^{i\mat \theta}, \\
	\mat D+\mat I=\text{sech}(\mat r)e^{i\mat \phi}.
\end{align}
where $\mat z=\mat re^{i\mat \theta}=e^{i \mat \theta^\tp}\mat r^\tp$. Therefore, numerically solving the equations \eqref{caponata}, \eqref{cefalina} with the initial conditions $\mat C(0)=\mat D(0)=\mat 0$ we can non-perturbatively solve the time evolution.

The covariance matrix $\mat \sigma$ of the detector+field state is defined as in the previous section. Once we have solved for the squeezing and rotation operators, in the sense that we have solved for $\mat z(t)$ and $\mat \phi(t)$, it is then straightforward to compute the evolved covariance matrix \cite{Heffner1}. If we split the covariance matrix into the block form
\begin{align}
	\mat \sigma=
	\begin{pmatrix}
		\mat \sigma_{qq} & \mat \sigma_{qp} \\
		\mat \sigma_{qp}^\tp & \mat \sigma_{pp}
	\end{pmatrix},
\end{align}
then straightforward algebraic operations (See Appendix \ref{appendixA}, Eqs.~\eqref{squeezeop1} and~\eqref{squeezeop2}) allow one to determine the form of these blocks. For example if our detector+field state starts in the vacuum state then the evolved state is simply given by a multi-mode squeezed state $\op{S}(\mat z)\ket{0}$, since the vacuum is invariant under rotations. In this case, the blocks of the evolved covariance matrix take the form
\begin{align} \label{covariance}
	\mat \sigma_{qq}=&\frac{1}{2}(\cosh(2\mat r)+\sinh(2\mat r)e^{i\mat \theta}+\cosh(2\mat r^\tp)\\
	&+\sinh(2\mat r^\tp)e^{-i\mat \theta^\tp}), \nonumber  \\
	\mat \sigma_{pp}=&\frac{1}{2}(\cosh(2\mat r)-\sinh(2\mat r)e^{i\mat \theta}+\cosh(2\mat r^\tp)\\*
	&-\sinh(2\mat r^\tp)e^{-i\mat \theta^\tp}), \nonumber  \\
	\mat \sigma_{qp}=&\frac{i}{2}(\cosh(2\mat r)-\sinh(2\mat r)e^{i\mat \theta}-\cosh(2\mat r^\tp)\\*
	&+\sinh(2\mat r^\tp)e^{-i\mat \theta^\tp}).\nonumber
\end{align}

Once this is obtained then computing the covariance matrix for the detector(s) alone is trivial: one must simply isolate the rows and columns of $\mat \sigma$ corresponding to the detector modes.  Both methods give the same time evolution and the same covariance matrix applied to the same scenarios, and we used them both independently to check our calculations.

\section{Results}
\label{sec:results}

As discussed in the introduction, we will see below how these tools can be used to observe relativistic quantum phenomena. After examining effects due to sharp switching \cite{Jorma1,Satz,Pad1} we will go on to study the emergence of causal signaling with increased number of field modes \cite{Fay}, the Unruh effect \cite{Unruh1, BandD} and vacuum entanglement harvesting \cite{Reznik1, VerSteeg2009}. We will be considering these effects in the context of optical cavities. Cavities are in fact excellent systems to study since experimental physicists have developed tools for the precise production and control of optical states inside cavities, superconducting SQUIDs, or microwave guides, and therefore such systems are likely to be key in the experimental verification of relativistic phenomena \cite{PastFutPRL}.

In all of the scenarios we consider here, we will simplify the calculations by using the detector's proper time \(\tau\) as our preferred time coordinate, as is done in the majority of the literature concerning Unruh-DeWitt detectors. Doing this is acceptable here because all of the cases we consider either involve only a single detector or two detectors that share the same proper time. In particular we will take for our interaction Hamiltonian the usual Unruh-DeWitt interaction, namely monopole-monopole coupling, but of course with the usual qubit operators replaced by their corresponding oscillator operators. 

Using the notation of section \ref{secrel}, in the interaction picture the interaction Hamiltonian is therefore
\begin{align} \label{UDWhamil}
	\op{H}^{\text{D},\tau}_{1}(\tau)
	&= \lambda(\tau)\sum_j (e^{-i \Omega_j \tau}\op{a}_{d_j}+e^{i\Omega_j \tau}\op{a}_{d_j}^\dag) \nonumber \\
	&\times \sum_n (u_n [x(\tau),t(\tau)]\op{a}_n+u_n^* [x(\tau),t(\tau)]\op{a}_n^\dag),
\end{align}
where $u_n$ is given by either Eq. (\ref{modefunction1}) or (\ref{modefunction2}), depending on the boundary conditions we impose. Since it is numerically simpler to change to the Heisenberg picture and use the equation of motion as presented in section \ref{phase}, we use this method for all calculations and use a second numerical method based on the formalism presented in section \ref{hilbert} as an independent check of our results.

As a note, it is important to realize that some of the following scenarios are actually solved exactly analytically (not numerically). We can do this in the cases where we have stationary detectors with constant (sharp) switching functions since in these cases the Hamiltonian commute with itself at different times. Equivalently, in these cases the solution to Eq. (\ref{step4}) is simply $\mat S(\tau)=\exp(\mat \Omega \mat F^{\text{sym}}\tau)$.

\subsection{Stimulation of noise due to sudden switching}

 Before examining more involved settings such as the effect of non-inertial motion on our detector or the vacuum entanglement harvesting, let us discuss first  what happens for a single inertial detector in a vacuum background. As is known, the use of sufficiently sharp switching functions $\lambda(\tau)$ can stimulate excitation of the detector even when it is inertial \cite{Pad1}. This stimulated vacuum fluctuations can be reduced by increasing the interaction time and using smooth switching functions, for example, Gaussian time profiles. Indeed,  we see non-negligible inertial excitation when the time of integration is small enough, but we will confirm that this vanishes for long duration Gaussian switching functions, as it corresponds to the detection of quantum noise in the vacuum state of a quantum field.
\begin{figure*}
	\centering
	         \includegraphics[width=0.45\textwidth]{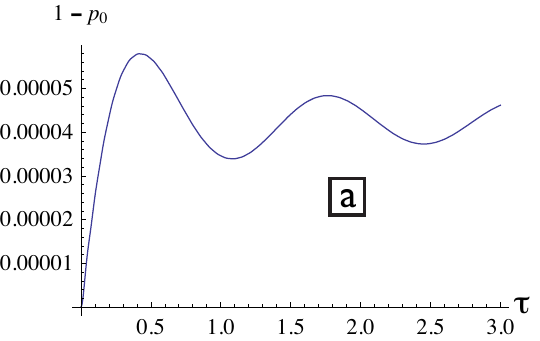}
	          \includegraphics[width=0.45\textwidth]{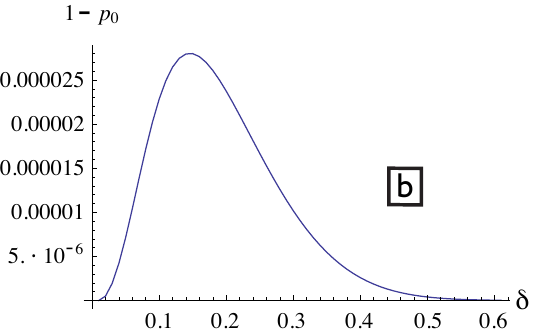}
	 \caption{Excitation probability of an inertial detector: In (a) we consider a sharp switching function that jumps suddenly from 0 to a constant value $\lambda$ and remains at that value thereafter. We consider the resulting final excitation probability of the detector if it were to be examined at proper time $\tau$ after being switched on (at $\tau=0$). Even when $\tau$ becomes much larger than the values shown in the plot, the probability remains nonzero. In (b) we consider Gaussian switching functions of standard deviation $\delta$, and we plot the excitation probability after the detector has finished evolving (i.e., when the Gaussian tail has become negligible); hence, $\delta$ in that plot is a measure of smoothness. For sharp switching, we observe excitation as expected, and for Gaussian switching, we observe excitation only when the Gaussian is sharp enough, which is also as expected. The parameters used are $\lambda=1/100$, $L=2\pi$, $\Omega=9/2$ (resonant with the ninth mode), and the detector is placed at position $x=L/2=\pi$.}
        \label{vacuumplots}
\end{figure*}
 
There have been some studies on the effect of the smoothness of the switching function on the probability of excitation of the Unruh-DeWitt detector and the stimulation of quantum noise \cite{Jorma1,Satz}. These studies showed that the quantum noise that a detector will observe for short interaction times is strongly influenced by the way in which the detector is switched on. These studies were done within perturbation theory and only for the case of fields in free space. It would be interesting to show these effects in cavity settings and in a non-perturbative calculation.

To this end, we consider a single detector sitting inertially in the center of a cavity of length $L$, such that $t(\tau)=\tau$ and $x(\tau)=L/2$. Seeing as we are considering an optical cavity, we will use reflecting boundary conditions for the field. We then solve for the evolution generated by the Hamiltonian \eqref{hamil2a} with some switching function $\lambda(\tau)$. After the evolution, our detector+field acquires  a multi-mode squeezed, pure-state covariance matrix $\mat \sigma$. The $2\times 2$ covariance matrix $\mat \sigma_d$ corresponding to the oscillator detector is then obtained by taking the detector-detector elements of $\mat \sigma$:
\begin{align}   \label{detector}
	\mat \sigma_d=
	\begin{pmatrix}
		\sigma^{(d)}_{qq} & \sigma^{(d)}_{qp} \\
		\sigma^{(d)}_{qp} & \sigma^{(d)}_{pp}
	\end{pmatrix}.
\end{align}
A useful quantity, the symplectic eigenvalue $\nu$ of this state, can easily be computed as the absolute value of either of the eigenvalues of the matrix $i\mat \Omega \mat \sigma_d$ (they come in a $\pm$ pair), where $\mat \Omega$ is the single-mode symplectic form
\begin{align}
	\mat \Omega=
	\begin{pmatrix}
		0 & 1 \\
		-1 & 0
	\end{pmatrix}.
\end{align}
The value of $\nu$ gives the mixedness of the state, with $\nu=1$ corresponding to a pure state. To be precise, the purity is given by $\text{Tr} \op\rho_d^2= \nu^{-1}$. In general it is somewhat nontrivial to compute the excitation probabilities $p_n=\bra{n}_d \op\rho_d \ket{n}_d$ of a given Gaussian state \cite{Dodonov1}, but for our purposes we will find it sufficient to only consider the probability of no excitation, $p_0$. In the case of a zero-mean state (i.e. the Gaussian Wigner function is centered at the origin of phase space), which includes the states we consider (as shown in the appendix \ref{appendixA}), this probability is given by  \cite{Dodonov1}
\begin{align} \label{p0}
	p_0=2/\sqrt{\text{det}\, \mat \sigma_d+ \text{Tr}\, \mat \sigma_d+1 }.
\end{align}

To examine the stimulated excitation of the detector due to the activation of the interaction we will consider two different switching functions. The first will represent sharp switching on and off, in which we set the switching function to a constant $\lambda(\tau)=\lambda$ for $\tau>0$ and zero otherwise, and track the evolution through time. In the second case we will use a Gaussian switching function of the form $\lambda(\tau)=\lambda \exp(-\tau^2/2\delta^2)$ and will integrate from time $\tau_i=-4\delta$ to $\tau_f=4\delta$. $\delta$ is in this case a measure of the smoothness of the time profile. Note that for sharp switching, considerably more field modes must be included before we observe solution convergence than in the case of Gaussian switching. This is expected since a sharp $\lambda(\tau)$ can excite field modes significantly higher than those near resonance. This is because the off-resonant rotating wave terms become important (as well as the counter-rotating wave terms) if the interaction suddenly changes in characteristic times faster than $\sim1/\Omega$; see \cite{Satz}. We then use Eq. (\ref{p0}) to compute the probability of excitation, $1-p_0$, for both the sharp switching as a function of $\tau$ and Gaussian switching as a function of $\delta$. The results are plotted in Fig. \ref{vacuumplots}.

In both cases we see that the excitation probability tends to zero as the interaction time goes to zero, as it must since this is the limit of no evolution. Note that these results can also be easily computed perturbatively, giving the same answer (up to higher than  second order corrections) as that shown in  Fig. \ref{vacuumplots}. For the Gaussian switching function we see that the excitation becomes negligible for larger $\delta$; this results from the switching function becoming smoother with increasing $\delta$ and is exactly what should be expected. For sharp switching however we see that the excitation probability does not decay with increasing time because in this case it is the initial discontinuity in $\lambda(\tau)$ at $\tau=0$ that causes the excitation.

\subsection{Emergence of Causal Signaling}\label{secsignaling}

The causal behaviour of the probability of excitation of spacelike separated atoms have been thoroughly studied in the continuum in the context of the so-called Fermi problem \cite{Fermi,Hegerfeld}, where satisfactory answers have been provided (see among others \cite{Sab1,Sab2}).

Regarding IR cutoff settings, it has been pointed out that a single mode approximation in the Unruh-Dewitt model (namely, considering a system of detectors interacting only with a single mode of the quantum field) leads to superluminal signaling \cite{Fay}. This is not surprising: a complete set of solutions to the field equations inside a cavity are the stationary waves inside it. A propagating signal cannot be expressed in terms of a finite number of stationary waves. Strictly speaking, to completely recover causality one should consider the infinite number of modes inside the cavity.

However, it is also known from quantum optics \cite{ScullyBook} that the Jaynes-Cummings model (basically a single-mode-approximated Unruh-DeWitt model) produces accurate results if the evolution times are long. This is so because for long times (much longer than the light crossing time of the cavity) a stationary regime is reached, and for infinite times there is, of course, no signal propagation issues. Hence, if we are to study quantum information-related topics with this model we need to take seriously the issue of causality. This is one reason why, for certain scenarios, the use of only a single mode is highly unphysical.

\begin{figure*}[t]
	\centering
                 \includegraphics[width=0.32\textwidth]{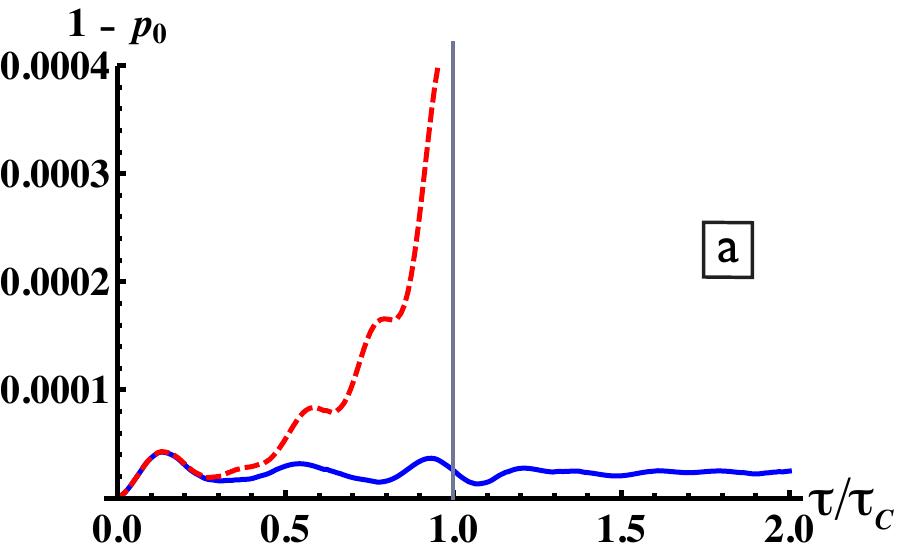}
                 \includegraphics[width=0.32\textwidth]{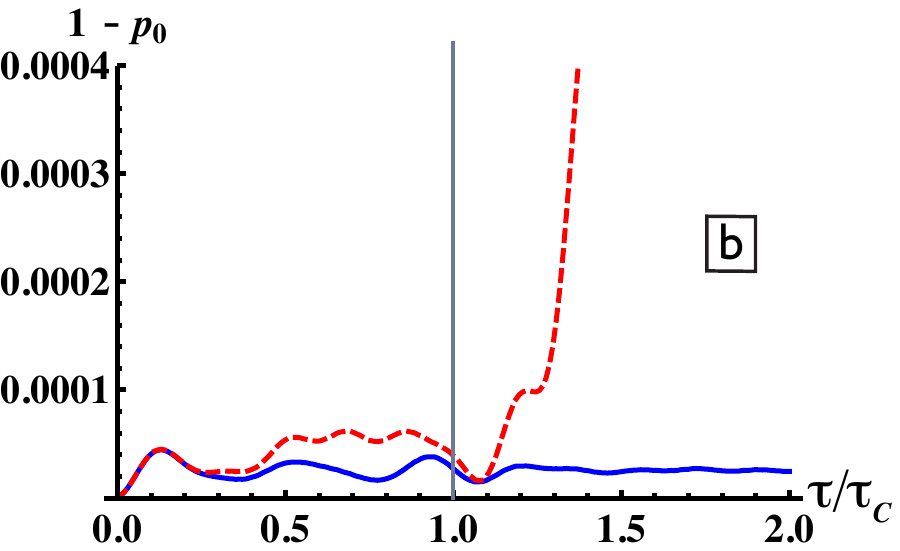}
                  \includegraphics[width=0.32\textwidth]{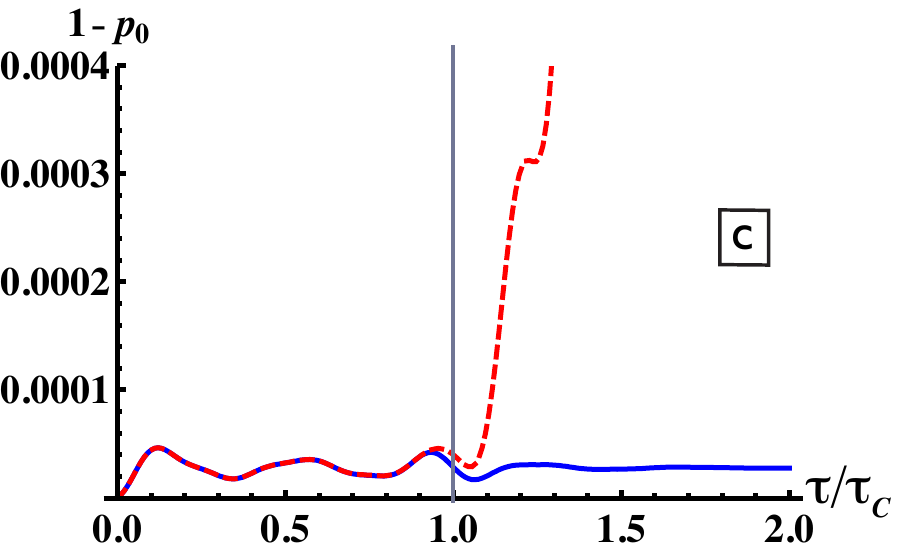}
	\caption{Excitation probability as a function of time of an inertial detector (detector~1) in the presence of another in a highly excited state (detector~2) considering (a) 10, (b) 13,  and (c) 16 field modes. Both detectors are sharply switched on at the same time, and the distance between them is such that the light-crossing time from one to the other is $\tau_C$. The vertical lines represent the time at which this mutual influence between detectors should becomes possible. In each plot we consider what happens when we include a different number of field modes in the calculation.  Detector~1 begins its evolution in its ground state. The solid (blue) lines represent the excitation probability of detector~1 when detector~2 is also started in its ground state, whereas the dashed (red) lines are when detector~2 is started in an excited state (squeezed with $r=5$). As expected, the initial squeezing of detector~2 contributes to the subsequent excitation of detector~1. However, this influence should not be able to reach detector~1 until $\tau/\tau_C=1$. By considering different numbers of field modes, we explicitly observe the emergence of causal signaling as the number of field modes is increased. The parameter values are  $L=2\pi$, $\lambda=1/100$ and $\Omega=9/2$ (resonant with the 9th mode).}
        \label{causalplots}
\end{figure*}

Here we point out the fact that, using our work, we can very easily observe the emergence of causal signaling by considering our detector to be coupled to different numbers of field modes. Although, rigorously speaking, we would need to add up the infinite number of modes to completely guarantee causality, we will discuss that, for a given arbitrarily high time precision, it is possible to find an effective model with a finite number of modes such that acausal signaling does not happen within the required time precision. To do so, we will consider two inertial detectors in the same cavity, and our goal is to examine how long it takes one detector to observe the effects of the other via propagating excitation in the field. We will choose sharp switching functions for both of the detectors. We will furthermore take the initial state of one of the detectors, say detector~2, to be highly excited, specifically a single mode squeezed state with a covariance matrix of the form
\begin{align}
	\mat \sigma_2=
	\begin{pmatrix}
		e^{2r} & 0 \\
		0 & e^{-2r}
	\end{pmatrix},
\end{align}
where $r$ is the squeezing parameter. Aside from this, we take the other detector and the field to be in their vacuum states. The entire initial detectors+field covariance matrix is therefore equal to the identity except for the two diagonal entries corresponding to detector~2. We then switch on both detectors at time $\tau=0$ and compute the excitation probability of detector~1 via Eq. (\ref{p0}).

We place the two detectors at positions $x_1=L/4$ and $x_2=3L/4$, such that they are a distance $L/2$ apart. Since we are taking the speed of light  $c=1$ this means that the time from the initial switching required for the detectors to come into causal contact is $\tau_c=L/2$. In Fig. \ref{causalplots} we display the excitation probability of detector~1 as a function of $\tau/\tau_c$, where we show the results including $10$, $13$ and $16$ field modes. 

The vertical line in Fig. \ref{causalplots} represents the moment in time $\tau=\tau_c$ at which the two detectors come into causal contact. In each of the plots the solid blue curve is the excitation probability for detector~1 when detector~2 is initially in its ground state $(r=0)$, whereas for the dashed (red) curve we initialize detector~2 in a squeezed state with squeezing parameter $r=5$. As expected, we observe increased excitations in detector~1 that are caused by the propagating field quanta emitted from the squeezed detector~2. We see however that if one does not include enough field modes, then the additional excitation occurs before the two detectors should be in causal contact, therefore implying superluminal signaling. It is only when we include enough field modes that we find the two curves diverging only at $\tau=\tau_c$, implying that this is when they have mutual influence. Note that, of course, even with 16 modes the signaling can be seen to be slightly acausal. Further increasing the number of modes further improves the causality, although it quickly becomes the case that one must include many more modes to see a slight improvement. It is only in the limit of infinite modes that we have causality in the strict sense.

As a final word on this, we point out that examining this emergence is easy using our method. First, the symplectic evolution $\mat S(\tau)$  can be solved {\it analytically} since we use stationary detectors with sharp switching functions. Second, in each individual plot from Fig. (\ref{causalplots}), the two different curves, solid and dashed, are computed using the same transformation $\mat S(\tau)$. The only difference is the initial state $\mat \sigma(0)$ that we evolve via $\mat \sigma(\tau)=\mat S(\tau)\mat \sigma(0)\mat S(\tau)^T$. Here we have only briefly examined the emergence of causal signaling---our formalism is very well suited for a more complete and encompassing study of the effect.

\subsection{Non-perturbative Detection of the Unruh Effect and Thermalization}\label{Uther}

In this section, we will use the formalism developed above to study, beyond perturbation theory, the Unruh effect inside a cavity. For a detector with uniform  acceleration $a$, the worldline $(t(\tau),x(\tau))$ to be used in our Hamiltonian (\ref{UDWhamil}) is given by
\begin{align}
	t(\tau)=a^{-1}\sinh(a\tau), \;\;\; x(\tau)=a^{-1}(\cosh(a\tau)-1),
\end{align}
such that the detector is at position $x=0$ at proper time $\tau=0$ \cite{Aasen}.

The excitation stimulated by insufficiently smooth switching functions can be somewhat of a problem in attempting to observe the Unruh effect in cavities. This is because for a given acceleration the crossing time of the detector from one side of the cavity to the other may be small enough to induce significant noise and wash out the Unruh noise. Increasing the acceleration (and therefore the Unruh noise) does not fix the problem because then the integration time is even less, and  that means  an increased amount of stimulated fluctuations. One way of overcoming this difficulty would be to simply increase the length $L$ of the cavity, however doing so also increases the number of significant modes that must be included in the integration and so greatly increases the computational effort required.

For this reason we have opted for a simpler solution. Namely, rather than using reflecting (Dirichlet) boundary conditions as would be the case in a linear cavity we will instead use periodic boundary conditions, as would be the case in a periodic waveguide \cite{PastFutPRL}. In this setting it is physically acceptable for us to arbitrarily increase the interaction time to values large enough that the stimulated noise becomes negligible and thus allowing a clearly observation of the Unruh effect. Although in taking this approach we are forced to include more modes (because we must now include the zero- and negative-frequency modes of the field), this is still more efficient than increasing the length of the cavity. Whatever method we utilize, considering long interaction times provides the desired results. Note that the problem can still be relatively challenging from the computational point of view since for extremely large accelerations the coupling matrices $\mat w(\tau)$ and $\mat g(\tau)$ from Eq.~\eqref{hamil2a} become highly oscillatory very quickly.

As an aside, there is further reason to use periodic boundary conditions instead of reflecting ones when studying accelerated detectors: In a reflecting cavity an accelerated detector will observe the field modes becoming increasingly blueshifted as it travels faster. Eventually even the fundamental mode will be shifted beyond the detector's resonance frequency, meaning that it no longer resonates with any of the modes, at which point the detector effectively decouples from the field. If we instead use periodic boundary conditions then this effective decoupling does not occur. To see why this is so, consider a detector accelerating to the right. According to this detector the left-moving modes of the field undergo an increasing blueshift as in the reflecting cavity. The right-moving modes however experience a red-shift, meaning that the detector resonates with higher right-moving modes over time rather than lower. Thus, as long as we include enough field modes in our calculation, the detector will continue experiencing resonance throughout its evolution. Using periodic boundary conditions we therefore avoid the blue-shift induced decoupling. This is better for studying the Unruh effect inside cavities since it is more similar to the physics of free space.

In order to test the Unruh effect with our model, we must ask two questions. First, after the interaction is complete, is our accelerated oscillator in a thermal state? Second, if so, does the temperature depend linearly on the acceleration? To answer the first question let us recall the definition of a thermal state in the Gaussian formalism \cite{Adesso1}. 
In the case of a single-mode thermal state, the covariance matrix is diagonal: $\mat \sigma_d^\text{therm}=\text{diag}(\nu,\nu)$, where $\nu$ is the state's symplectic eigenvalue. The excitation probabilities in this case follow a Boltzmann distribution:
\begin{align}  \label{ptherm}
	p_n^\text{therm}=\frac{2}{\nu+1}\left(\frac{\nu-1}{\nu+1} \right)^n,
\end{align}
with corresponding temperature
\begin{align}  \label{temp}
	T=\Omega\left[ \ln \left(1+\frac{2}{\nu-1} \right)\right]^{-1}.
\end{align}

As explained in appendix \ref{appendixA} the time evolution generated by Eq. (\ref{UDWhamil}) is given by a multi-mode squeezing unitary $\op{S}$. The means that the detector+field state after evolution is a multi-mode squeezed state of the form  $\op{S}\ket{0}$. It is known that the reduced state corresponding to a subsystem of a multi-mode squeezed state is not in general given by a thermal state. Rather, it is given by a squeezed thermal state \cite{Schumaker1}. This is because the squeezing operation $\op{S}$ generally includes both inter-mode squeezing (which produces thermal subsystems) as well as single-mode squeezing. 

The following question then arises: in our specific scenario, does the detector evolve into purely a thermal state, or has it also been squeezed? To answer this we compute the symplectic eigenvalue $\nu$ of our detector state $\mat \sigma_d$, as given by the absolute value of either of the eigenvalues of the matrix $i\mat \Omega \mat \sigma_d$, and compare the probability spectrum of this state with that of a thermal state that is given the same symplectic eigenvalue \cite{Adesso1}. From the cases that we have examined, it appears the answer is that while the detector does not become exactly thermal, it is very nearly so. That is, the multimode squeezing between the detector and the field modes is much greater than the single-mode squeeze undergone by the detector. To conclude this, we computed the probability of no excitation from Eq. (\ref{p0}), as well as $p_0^\text{therm}$ and $p_1^\text{therm}$ from Eq. (\ref{ptherm}), i.e. what the probabilities of zero and first excitation would be were our state a thermal state with the same symplectic eigenvalue. For small temperatures and therefore small excitation probability, which is what we will consider here, a good test for thermality is whether or not we have $p_0-p_0^\text{therm} \ll p_1^\text{therm}$. If this is satisfied, then the detector is very nearly thermal. For the cases we have examined, we have found that $p_0-p_0^\text{therm}$ is approximately seven orders of magnitude less than $ p_1^\text{therm}$, and we therefore conclude that our model does indeed produce the detector thermality expected. With this first question answered, we are free to examine the temperature dependence on acceleration, where the temperature is given by Eq. (\ref{temp}). 

In the following example, we start with a negative initial value for the proper time, $\tau_i<0$, and evolve to the positive time $\tau_f=-\tau_i$. This lets us use a long interaction time while minimizing the computational effort. That is, we imagine that the detector is injected at high velocity into our waveguide such that the acceleration is in the opposite direction to its motion. The detector then slows down, comes to a stop at $\tau=0$, and then begins looping around in the other direction before exiting at the same speed it entered with. Again, we are using periodic boundary conditions for the field so that this setup makes physical sense. We use a Gaussian switching function $\lambda(\tau)=\lambda \exp(-\tau^2/2\delta^2)$ with $\delta$ large enough that the switching stimulation is negligible. It is after this evolution is finished that we compute the temperature of the detector.

\begin{figure}[t]
        \includegraphics[width=0.45\textwidth]{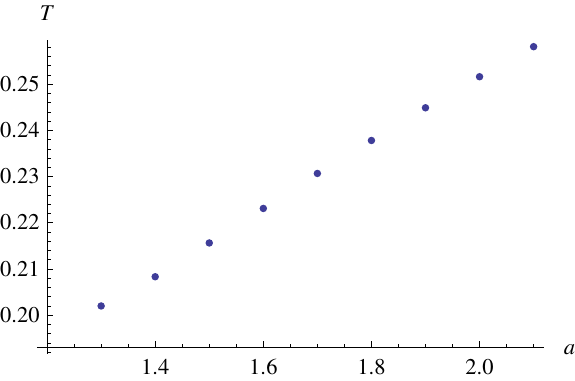}
	\caption{Temperature (after evolution) of an accelerated detector as a function of its acceleration. We  observe a linear dependence as expected from the Unruh effect. The parameters used were $L=4\pi$, $\lambda(\tau)=\lambda \exp(-\tau^2/2\delta^2)$ with $\lambda=1/100$ and $\delta=8/7$, and the detector gap was $\Omega=4$ (resonant with the 8th field modes).}
\label{Tvsa}
\end{figure}

For a given set of parameters (see caption), we plot in Fig. (\ref{Tvsa}) the temperature of the detector as a function of its acceleration.
We see that the detector temperature indeed depends linearly on its acceleration, in concurrence with the prediction of the Unruh effect. This is excellent since the question of the response of an accelerated detector inside a cavity has previously been unexplored and indeed has generated some debate in discussion. Our result represents, to the knowledge of the authors, the first confirmation of the Unruh effect occurring inside of a cavity and, moreover, in a nonperturbative fashion.

One may worry that extrapolating our data backwards it seems that the temperature does not vanish at $a=0$. This is due to a couple of factors. First, since we are in a cavity we should not expect the Unruh effect to hold for very low accelerations. For very low accelerations, the characteristic length $c^2/a$ of the acceleration will be much greater than the length of the cavity, at which point we expect to see significant border effects, so the response of the detector will not necessarily be thermal. Second, as one goes to very low temperature, the corresponding probability of excitation is exponentially suppressed. This means that for very small temperatures the Unruh effect will be washed out by the switching noise even if the switching function is very smooth.

\subsection{Vacuum Entanglement Harvesting}

In addition to the Unruh effect, another relativistic quantum phenomenon that is of great interest is the extraction of entanglement from the vacuum field. That is, two detectors can become entangled by each interacting locally with a quantum field, even if they remain spacelike separated \cite{Reznik1, VerSteeg2009,Sab2,PastFutPRL}. Of course, it is well known that no local operations can increase entanglement between two quantum systems \cite{mikeandike}. In the case at hand however there is already entanglement present in the vacuum state of the field between spatially separated degrees of freedom, and so by interacting with the field locally, multiple detectors can extract this entanglement to become entangled themselves. This is true even if the detectors remain spacelike separated throughout their evolution, meaning that they can become entangled without any direct causal mutual influence.

In our detector model, we can consider multiple detectors very easily by simply adding their respective field interactions into the coupling matrices $\mat w$ and $\mat g$ used in Eq. (\ref{hamil2a}). Once the evolution has been solved and the detectors+field covariance matrix $\mat \sigma$ obtained, the multi-detector covariance matrix $\mat \sigma_d$ is obtained by deleting the rows and columns corresponding to the field. In the case of two detectors, we obtain a $4\times 4$ covariance matrix that can be arranged in the form
\begin{align}
	\mat \sigma_d=
	\begin{pmatrix}
		\mat \sigma_1 & \mat \sigma_{12} \\
		\mat \sigma^\tp_{12} & \mat \sigma_2
	\end{pmatrix},
\end{align}
where $\mat \sigma_1$ and $\mat \sigma_2$ are the $2\times 2$ covariance matrices of the detector-1 and detector-2 subsystems, and are of the same form as Eq. (\ref{detector}). $\mat \sigma_{12}$ is a $2\times 2$ matrix that provides the correlations between the two detectors. For a two-mode Gaussian state such as this, the entanglement between the detectors, as measured by the logarithmic negativity, is given by \cite{Adesso1}
\begin{align}
	E_N=\text{max}(0,-\log \nu_-^\Gamma),
\end{align}
where $\nu_\pm^\Gamma=\sqrt{(\Delta \pm \sqrt{\Delta^2-4 \text{det}\mat \sigma_d})/2}$ are the symplectic eigenvalues of the partially-transposed covariance matrix, and $\Delta=\text{det} \mat \sigma_1+ \text{det} \mat \sigma_2-2\text{det} \mat \sigma_{12}$.
\begin{figure}[t]
	\centering
        \includegraphics[width=0.45\textwidth]{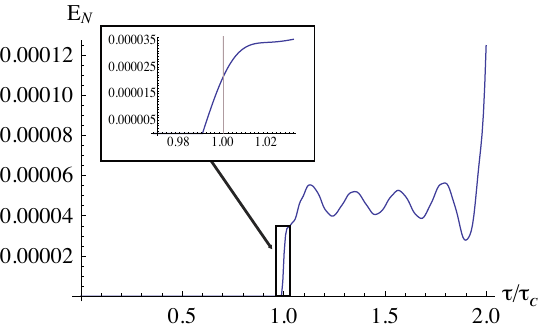}
	\caption{Extraction of vacuum entanglement with two detectors. Notice that the entanglement harvesting starts close to the time $\tau/\tau_c=1$, where the detectors come into causal contact but still before they reach this time. The parameters used are $L=2\pi$, $\lambda=1/100$ and $\Omega=9$ (resonant with the 18th mode).}
        \label{lognegplot}
\end{figure}

Using this, we plot in Fig. (\ref{lognegplot}) an example of the logarithmic negativity between two detectors as a function of $\tau/\tau_C$, where $\tau_C$ is the time it takes for the detectors to come into causal contact. We used sharp switching functions for both detectors, as we observed that the stimulation incurred by switching aided in generating entanglement. Note that due to this sharp switching, we needed to include many modes, up to $100$, in order to find convergence of our solution. The positions of the detectors in this case were chosen to be $x_1=L/4$ and $x_2=3L/4$, implying that $\tau_C=L/2$. We indeed observe that, given enough time to locally interact with the field, the two detectors become entangled. Furthermore, we see in this example that entanglement is produced before the detectors come into causal contact, although just barely.

As a final note, if we compare a harmonic-oscillator particle detector with the standard qubit detector, one can easily show that to second order in perturbation theory (leading order in this phenomenon) the only difference between the evolution of a single qubit and that of a single oscillator is that the oscillator develops off-diagonal coherence terms in its density matrix. In view of this fact, one might hypothesize that oscillators are less efficient at extracting spacelike entanglement than are qubits, which would make two-level systems more appropriate for analysis of field entanglement harvesting. This remains an open question.

\section{Conclusions}

By applying the Gaussian formalism, we have addressed the problem of time evolution of a particle detector undergoing relativistic movement inside of a cavity. With this, we are able to tackle arbitrary multimode time-dependent problems and solve them nonperturbatively. This is markedly different from the standard Unruh-DeWitt model that can generally only be solved perturbatively. Remarkably, the only fundamental change between the standard approach and our work is that we use a harmonic oscillator to describe a detector, rather than a qubit.

In addition to being nonperturbative, the methods we have presented lead to a computationally efficient way of solving a great range of problems involving an arbitrary number particle detectors coupled to quantum fields inside a cavity. The flexibility of the model extends to the following: (1) the detectors can undergo arbitrary relativistic motion; (2) they can have arbitrary quadratic interaction with the field; (3) the field and detectors can begin in any Gaussian initial state; and (4) our description of the field can include any number of modes with arbitrary time-dependent boundary conditions. The vast range of scenarios that this can encompass are all solved by the same number-valued, linear, first-order ordinary differential equation. We have the additional advantage that for a given evolution, we do not need to solve the equation again if we decide to change the initial state. To demonstrate this wide applicability, we have analyzed a range of different problems of interest in general relativistic quantum field theory, obtaining several results.

One of our most important findings is that an accelerated harmonic-oscillator detector in a cavity exhibits a thermal response with temperature proportional to its acceleration. Namely, we have demonstrated that the Unruh effect occurs inside cavities, a scenario that we believe has not been previously explored. We emphasize that we obtain the thermal response from the movement of the detector, as opposed to assuming a different quantization scheme for accelerated observers and going through Bogoliubov transformations. We thereby show evidence pointing towards the universality of the Unruh effect: a thermal response proportional to the detectorŐs acceleration appears even (1) considering a different model instead of a standard Unruh-DeWitt detector, (2) inside a cavity, and (3) nonperturbatively. Also, we are able to analyze the border effects appearing due to the finite size of the cavity.

We have been able to quantify the effects of sudden switching for inertial detectors, determining the excitation probability in different scenarios. The results obtained are as expected: the sharper the switching function, the greater the stimulation generated.
Remarkably, we also quantitatively address the problem of recovering causality in the context of fully relativistic cavity field theory in which there is a UV cutoff. We have shown that whereas a single-mode model strongly violates causality, causal signaling emerges as more modes in the field are included. Although rigorously speaking, an infinite number of discrete modes are required to recover causality, we can determine how many modes are necessary to include in the problem to have causal signaling for a given time precision.

We also analyzed the problem of vacuum entanglement harvesting. We have shown that indeed, harmonic-oscillator detectors inside a cavity can extract entanglement from the vacuum field and can achieve this entanglement while they are spacelike separated.

We close emphasizing again the immense applicability of our model. There are many problems of interest that can now be easily addressed, and we hope that our work will be used in the future to explore the problems of relativistic quantum physics. For example, these methods are well suited to studying the Unruh effect in a variety of models since it is straightforward to modify the detector-field interaction and to consider arbitrary detector trajectories in flat spacetime. Applications to cavity settings in curved spacetimes can be also considered. For instance, it could be easily shown how to translate some of the results to cavities close to the event horizon of a stationary black hole \cite{Kerr},  and  it would be undoubtedly interesting to analyze what would be the behaviour of these quantum systems when they experience a dynamical gravitational collapse  \cite{colapse,Collapse2}.

{\bf Note added}: During the final stages of this project, we became aware of another research group working along similar lines ~\cite{NottinghamGroup}. We both agreed to release our completed work simultaneously.

\acknowledgments 
We are grateful to Achim Kempf for  invaluable discussions  during the development of this work. We also thank Ivette Fuentes for pointing out previous literature on the subject and Tim Ralph for hosting the RQI6 conference, which provided valuable feedback and an opportunity to fine-tune our work.  N.C.M.\ is grateful to The University of Waterloo Department of Applied Mathematics, Department of Physics and Astronomy, and Institute for Quantum Computing for hospitality during several visits that provided crucial collaboration opportunities for this project. E.M.-M.\ and E.G.B.\ similarly thank The University of Sydney School of Physics for hospitality during their visit. This work was supported in part by the Natural Sciences and Engineering Research Council of Canada. E.M.-M.\ acknowledges the support of the NSERC Banting Postdoctoral Fellowship programme. N.C.M.\ is supported by ARC Discovery Early Career Researcher Award DE120102204, which also provided partial support for E.M.-M.'s and E.G.B.'s visit to Sydney.

\appendix

\section{Detailed derivation of the calculation of the Hilbert space evolution}
\label{appendixA}


We will provide here the full detail of the derivation of the results presented in  section \ref{hilbert}.

As said in the main text, for a quadratic Hamiltonian, time evolution can be expressed in terms of displacements, squeezing, and rotation unitary operations. In particular, the unitary evolution we are looking to solve for can then be put into the form
\begin{align} \label{combo}
	\op U(\tau)=e^{i\gamma(\tau)}\op S(\mat z(\tau))\op D(\vec \beta(\tau))\op R(\mat \phi(\tau))\,,
\end{align}
where $\gamma$ is number valued and the squeezing, rotation and displacement operators are respectively defined as \cite{Ma1}
\begin{align} \label{squeeze}
	\op S(\mat z) &= e^{ \frac{1}{2} \big[(\opvec a^\dag)^\tp  \mat z \opvec a^\dag- \opvec a^\tp\mat z^\text{H} {\opvec a}\big] }, \\ \label{rotation}
	\op R(\mat \phi)&= e^{i (\opvec a^\dag)^\tp {\mat \phi} {\opvec a}}, \\ 
	\op D(\vec \beta)&= e^{\vec \beta^\tp \opvec a^\dag -\vec \beta^\herm \opvec a},
\end{align}
where $\mat z$ and $\mat \phi$ are matrices, and $\vec \beta$ is a column vector. The notation $\mat z^\text{H}$ is used to represent the conjugate transpose of $\mat z$, the elements of which are number valued. Note that we without loss of generality we can consider $\mat z$ to be symmetric because it is only the symmetric part that contributes to $\op{S}(\mat z)$. Also note that $\mat \phi$ must be a Hermitian matrix ($\vec \phi=\vec\phi^{\text{H}}$) to ensure unitarity of $\op{R}(\mat \phi)$.

The exact form of these transformations can be obtained nonperturbatively by employing a technique introduced by Heffner and Louisell~\cite{Heffner1}. We will utilize the polar decomposition of $\mat z$ into a product of a hermitian and a unitary matrix, which can always be achieved. This takes the form
\begin{align}
	\mat z=\mat re^{i \mat \theta}=e^{i\mat \theta^\tp}\mat r^\tp\,,
\end{align}
where $\mat r$ and $\mat \theta$ are hermitian matrices, and the second equality results from the assumed symmetry of $\mat z$. From here we wish to evaluate how such operators evolve the ladder operators of our system so that we can determine their corresponding symplectic transformations on the phase space and therefore the covariance matrix ascribed to, for example, a multimode squeezed state. Using the BCH's Hadamard lemma, $e^A B e^{-A}=B+[A,B]+[A,[A,B]]/2!+ \dots$, it is straightforward to obtain
\begin{align}
	&\op{S}^\dag (\mat z) \opvec a \op{S}(\mat z)=\cosh(\mat r) \opvec a+\sinh(\mat r)e^{i\mat \theta}\opvec a^\dag, \label{squeezeop1}  \\
	&\op{R}^\dag (\mat \phi) \opvec a \op{R}(\mat \phi)=e^{i\mat \phi}\opvec a, \\
	&\op{D}^\dag (\vec \beta) \opvec a\op{D}(\vec \beta)=\opvec a+\vec \beta,
\end{align}
and similarly
\begin{align}
	&\op{S}^\dag (\mat z) \opvec a^\dag \op{S}(\mat z)=\cosh(\mat r^\tp) \opvec a^\dag+\sinh(\mat r^\tp)e^{-i\mat \theta^\tp}\opvec a, \label{squeezeop2}  \\
	&\op{R}^\dag (\mat \phi) \opvec a^\dag \op{R}(\mat \phi)=e^{-i\mat \phi^\tp}\opvec a^\dag, \\
	&\op{D}^\dag (\vec \beta) \opvec a^\dag \op{D}(\vec \beta)=\opvec a^\dag+\vec \beta^*,
\end{align}

For our purposes, throughout the rest of this appendix we will ignore the contribution from the displacement operator. This is because we typically interested here in Hamiltonians that are quadratic and without linear terms. Since linear terms are what drive displacements, we need not consider them here. Generalizing to include displacements is however quite straightforward.

We will consider an interaction Hamiltonian in the interaction picture that takes the form
\begin{align} \label{interactionHamiltonian}
	\hat{H}^\text{D}_1= (\opvec a^\dag)^\tp \mat w(\tau) \opvec a +(\opvec a^\dag)^\tp \mat g(\tau) \opvec a^{\dag} +\opvec a^\tp \mat g(\tau)^\herm \opvec a\,
\end{align}
and we wish to solve for the unitary evolution $\op{U}(\opvec a^\dag,\opvec a,\tau)$ that it generates.

Consider now this unitary operator in its normal-ordered form $\op{U}^{(n)}(\opvec a^\dag,\opvec a,\tau)=\op{U}(\opvec a^\dag,\opvec a,\tau)$, where, for example, $(\op{a}_d \op{a}_d^\dag)^{(n)}=\op{a}_d^\dag \op{a}_d+1$. As explained in \cite{Heffner1}, we can equally well represent this using a number-valued function corresponding to $\op{U}^{(n)}$ of the form $\op{U}^{(n)}(\opvec a^\dag,\opvec a,\tau) \rightarrow \bar{U}^{(n)}(\vec \alpha^*,\vec \alpha,\tau)$, where $\vec \alpha$ and $\vec \alpha^*$ are taken to be column vectors consisting of real, independent variables. That is, we put $\op{U}$ into normal form and replace $\opvec a$ and $\opvec a^\dag$ by vectors of number-valued entries $\mat \alpha$ and $\vec \alpha^*$. In this representation, Schr$\ddot{\text{o}}$dinger's equation $i \partial_\tau \op{U}(\tau)=H_I (\opvec a^\dag, \opvec a, \tau)\op{U}(\tau)$ becomes
\begin{align} \label{schrod2}
	i\frac{\partial}{\partial \tau}\bar{U}^{(n)}(\vec \alpha^*\!\!,\vec \alpha,\tau)\!=\!\bar{H}_I^{(n)}\!\left(\vec \alpha^*\!\!, \vec \alpha+\frac{\partial}{\partial \vec \alpha^*},\tau \right)\bar{U}^{(n)}(\vec \alpha^*\!\!,\vec \alpha,\tau),
\end{align}
where $\bar{H}_I^{(n)}(\vec \alpha^*,\vec \alpha+\partial/\partial \vec \alpha^*,\tau)$ is obtained by putting $\op{H}_I$ into normal-ordered form (which in our case it already is) and replacing $\opvec a$ and $\opvec a^\dag$ by $\vec \alpha+\partial/\partial \vec \alpha^*$ and $\vec \alpha^*$ respectively. What we now have is a set of coupled, ordinary differential equations. An ansatz for the solution that we will use is $\bar{U}^{(n)}=e^{G(\vec \alpha^*,\vec \alpha,\tau)}$, turning the equation into one for $G$. Once the solution has been found, we can then obtain the normal ordered unitary by replacing back $\vec \alpha$ and $\vec \alpha^*$ by $\opvec a$ and $\opvec a^\dag$ and applying the normal ordering operator $\op{U}^{(n)}=\lcolon e^{G(\opvec a^\dag,\opvec a,\tau)} \rcolon$ where, for example, $\lcolon \op a_d \op a_d^\dag \rcolon=\op a_d^\dag \op a_d$. 

Following the prescription of \cite{Heffner1}, we now want to find the evolution equation of the number-valued function $\bar{U}^{(n)}$. From Eq. (\ref{schrod2}), we have
\begin{align}
	i\frac{\partial \bar{U}^{(n)}}{\partial \tau }=\bigg[ (\vec \alpha^*)^\tp \mat w \left(\vec \alpha+\frac{\partial}{\partial \vec \alpha^*}\right)+(\vec \alpha^*)^\tp \mat g \vec \alpha^* \nonumber \\
	 +\left(\vec \alpha^\tp+\frac{\partial}{\partial (\vec \alpha^*)^\tp}\right)\mat g^\text{H} \left(\vec \alpha +\frac{\partial}{\partial \vec \alpha^*}\right) \bigg] \bar{U}^{(n)},
\end{align}
and making the ansatz $\bar{U}^{(n)}=e^G$, we have the equation for $G$:
\begin{align} \label{Geqn}
	i \frac{\partial G}{\partial \tau}&= (\vec \alpha^*)^\tp \mat w \vec \alpha + (\vec \alpha^*)^\tp \mat w\frac{\partial G}{\partial (\vec \alpha^*)^\tp}
+ (\vec \alpha^*)^\tp \mat g \vec \alpha^* \nonumber \\
&+\vec \alpha^\tp \mat g^\text{H} \vec \alpha 
+\vec \alpha^\tp \mat g^\text{H} \frac{\partial G}{\partial \vec \alpha^*}
+\frac{\partial G}{\partial (\vec \alpha^*)^\tp}\mat g^\text{H} \vec \alpha \nonumber \\
&+ \frac{\partial G}{\partial (\vec \alpha^*)^\tp}\mat g^\text{H} \frac{\partial G}{\partial \vec \alpha^*}
+\frac{\partial}{\partial (\vec \alpha^*)^\tp}\mat g^\text{H} \frac{\partial G}{\partial \vec \alpha^*}. 
\end{align}
Additionally, we can make the educated ansatz 
\begin{align} \label{ansatz}
	G= (\vec \alpha^*)^\tp  \mat D(\tau) \vec \alpha + (\vec \alpha^*)^\tp \mat C(\tau) \vec \alpha^*+\vec \alpha^\tp \mat F(\tau) \vec \alpha+A(\tau),
\end{align}
where $\mat D$, $\mat C$ and $\mat F$ are matrices. In general, we should also include terms linear in $\mat \alpha$ and $\vec \alpha^*$, corresponding to phase space displacements, but in our case they will be absent due to the lack of linear terms in the relevant Hamiltonians and so we will not consider them. From here it is easy to show that
\begin{align}
	\frac{\partial G}{\partial \vec \alpha^*}=\mat D\vec \alpha+2\mat C_s \vec \alpha^*,
\end{align}
where $\mat C_s=(\mat C+\mat C^\tp)/2$ is the symmetric part of $\mat C$. The transposed version of this relation follows trivially.
Lastly, it is easily shown that
\begin{align}
	\frac{\partial}{\partial (\vec \alpha^*)^\tp}\mat g^\text{H} \frac{\partial G}{\partial \vec \alpha^*}=2 \text{Tr}(\mat g^\text{H} \mat C_s).
\end{align}

Given these relations it is now a simple matter of comparing coefficients between the right and left sides of Eq. (\ref{Geqn}). Doing so, we find the coupled set of differential equations
\begin{align} \label{Aeqn}
	&i\dot{A}=2 \text{Tr}(\mat g^\text{H} \mat C_s), \\ \label{Ceqn}
	&i\dot{\mat C}=4\mat C_s \mat g^\text{H} \mat C_s+2\mat w \mat C_s+\mat g, \\  \label{Deqn}
	&i \dot{\mat D}=(4\mat C_s \mat g^\text{H}+\mat w)(\mat D+\mat I), \\   \label{Feqn}
	&i\dot{\mat F}=(\mat D^\tp+\mat I)\mat g^\text{H} (\mat D+\mat I),
\end{align}
where $\mat I$ is the identity matrix, and we have initial conditions $A(0)=0$ and $\mat C(0)=\mat D(0)=\mat F(0)=\mat 0$.

These equations can be numerically solved, although we will find that for our purposes the only one that actually needs to be solved is the equation for $\mat C$. This is because $\mat C$ fully determines the squeezing matrix $\mat z=\mat re^{i\mat \theta}$, which, since our system is initially in the vacuum state, is all that we need (since the vacuum is invariant under rotations). That is why we can thus ignore the rotation and effectively set $\mat  \phi=\mat 0$. For a more general initial state, one would need to additionally solve for $\mat D$ in order to compute $\mat \phi$. Note also that one will never have to solve Eq. (\ref{Feqn}) for $\mat F$; it can be expressed purely in terms of $\mat C$ and $\mat D$ and is therefore a redundant variable.

Note that the form of these equations are entirely independent of the specific coupling matrices $\mat w$ and $\mat g$ that we choose. We are therefore free to choose an entirely different interaction Hamiltonian, and the evolution will still be represented by these equations. Once solutions have been found, we can return the (normal ordered) unitary to its operator form via $\op{U}^{(n)}=\lcolon e^{G(\opvec a^\dag,\mat{ \op{a}})} \rcolon$, which from Eq. (\ref{ansatz}) gives us
\begin{align} \label{soln1}
	\op{U}^{(n)}(\tau)=e^{A(\tau)}e^{(\opvec a^\dag)^\tp \mat C(\tau) \opvec a^\dag}\lcolon e^{(\opvec a^\dag)^\tp \mat D(\tau) \opvec a} \rcolon e^{\opvec a^\tp \mat F(\tau)\opvec a}.
\end{align}

Our problem is thus essentially solved; the last task required is to overcome the normal ordering and to put this unitary into a form that we are familiar with. The procedure for doing this is given in \cite{Ma1}, but we will reiterate it here in somewhat more detail. We know that $\op{U}$ should be a product of rotation and squeezing operators, along with possible phases:
\begin{align} \label{soln2}
	\op{U}=e^{i\gamma}\op{S}(\mat z)\op{R}(\mat \phi),
\end{align}
where $\op S$ and $\op R$ are given by Eqs. (\ref{squeeze},\ref{rotation}). Note that in our case $\op U$ contains no displacement operator because $\op{H}_{1}^{\text{D}}$ contains no linear terms. We know that the solutions Eqs. (\ref{soln1}) and (\ref{soln2}) must be equivalent, and the task is now to find the relation between $\gamma$, $\mat z$, and $\mat \phi$ and the results obtained for $A$, $\mat C$, $\mat D$ and $\mat F$. Trivially, we see that $i\gamma=A$ and from Eq. (\ref{Aeqn}), however this merely contributes an overall phase to the evolution, and we will therefore ignore this contribution henceforth since it does not contribute to the physics

In order to find $\mat z=\mat re^{i\mat \theta}=e^{i \mat \theta^\tp}\mat r^\tp$ and $\mat \phi$ we recall the action that Eq. (\ref{soln2}) will have on the ladder operator:
\begin{align} \label{action1}
	&\op{U}^\dag \opvec a \op{U}=\cosh(\mat r)e^{i\mat \phi}\opvec a +\sinh(\mat r)e^{i\mat \theta}e^{-i\mat \phi^\tp}\opvec a^\dag, \\ \label{action2}
	&\op{U}^\dag \opvec a^\dag \op{U}= \cosh(\mat r^\tp)e^{-i \mat \phi^\tp}\opvec a^\dag + \sinh(\mat r^\tp)e^{-i \mat \theta^\tp}e^{i\mat \phi}\opvec a.
\end{align}
We now need to compute $\op{U}^\dag \opvec a \op{U}$ and $\op{U}^\dag \opvec a^\dag \op{U}$ from the unitary in Eq. (\ref{soln1}) in order to compare. To this end, we use the identities $[\opvec a,\op{U}]=\partial \op{U}/\partial \opvec a^\dag$ and $[\opvec a^\dag,\op{U}]=-\partial \op{U}/\partial \opvec a$, or equivalently
\begin{align} \label{aidentities}
	\op{U}^\dag \opvec a \op{U}=\op{U}^\dag \frac{\partial \op{U}}{\partial \opvec a^\dag}+\opvec a, \;\;\; \op{U}^\dag \opvec a^\dag \op{U}=-\op{U}^\dag \frac{\partial \op{U}}{\partial \opvec a}+\opvec a^\dag
\end{align}
To evaluate the right-hand sides, we use the identities:
\begin{align}
	&\frac{\partial}{\partial \opvec a} \lcolon e^{(\opvec a^\dag)^\tp \mat D \opvec a} \rcolon = \mat D^\tp \opvec a^\dag   \lcolon e^{(\opvec a^\dag)^\tp \mat D \opvec a} \rcolon, \\
	&\frac{\partial}{\partial \opvec a^\dag}  \lcolon e^{(\opvec a^\dag)^\tp \mat D \opvec a} \rcolon=  \lcolon e^{(\opvec a^\dag)^\tp \mat D \opvec a} \rcolon \mat D\opvec a, \\
	&(I+\mat D^\tp)\opvec a^\dag   \lcolon e^{(\opvec a^\dag)^\tp \mat D \opvec a} \rcolon=  \lcolon e^{(\opvec a^\dag)^\tp \mat D \opvec a} \rcolon \opvec a^\dag, \\
	&\opvec a^\dag e^{\opvec a^\tp \mat F \opvec a}=e^{\opvec a^\tp \mat F \opvec a} (\opvec a^\dag-2\mat F\opvec a),
\end{align}
along with the fact that $\mat F$ is symmetric, as can be seen from Eq. (\ref{Feqn}). With these, Eq. (\ref{aidentities}) gives us
\begin{align}
	&\op{U}^\dag \opvec a \op{U}=[(\mat D+\mat I)-4\mat C_s (\mat D^\tp+\mat I)^{-1}\mat F]\opvec a \nonumber \\
	&+2\mat C_s(\mat D^\tp+\mat I)^{-1}\opvec a^\dag, \\
	&\op{U}^\dag \opvec a^\dag \op{U}=(\mat D^\tp+\mat I)^{-1}(-2\mat F \opvec a+\opvec a^\dag).
\end{align}
Since these equations are just the adjoints of each other, we are able to compare the two and determine the additional relations
\begin{align}
	\mat F=-(\mat D^*+\mat I)^{-1}\mat C_s^*(\mat D+\mat I), \\
	\mat I-4\mat C_s \mat C_s^*=(\mat D+\mat I)(\mat D^\dag+\mat I).
\end{align}
Given all of this, we find indeed that $\op{U}^\dag \opvec a \op{U}$ and $\op{U}^\dag \opvec a^\dag \op{U}$ are of the form given in Eqs. (\ref{action1},\ref{action2}), where we identify
\begin{align}
	\mat C_s=\frac{1}{2}\tanh(\mat r)e^{i\mat \theta}, \\
	\mat D+\mat I=\text{sech}(\mat r)e^{i\mat \phi}.
\end{align}
Thus, once we have integrated Eqs. (\ref{Ceqn},\ref{Deqn}) for $\mat C$ and $\mat D$ we can use this result to solve for the corresponding squeezing and rotation matrices $\mat r$, $\mat \theta$, and $\mat \phi$ and, via Eq. (\ref{covariance}), obtain the covariance matrix in which all properties of the final state are encoded.

\bibliography{RQIP}

\end{document}